\documentclass[reprint,
superscriptaddress,
amsmath,amssymb,
apl,
]{revtex4-2}
\usepackage[utf8]
{inputenc}
\usepackage{float}
\usepackage{titletoc}
\usepackage[usenames,dvipsnames]{color}
\usepackage{graphicx}
\usepackage{siunitx}
\usepackage{adjustbox}
\graphicspath{{./images/} }

\usepackage{textcomp}
\usepackage{dcolumn}
\usepackage{bm}
\usepackage{comment}         
\usepackage[usenames,dvipsnames]{color}
\usepackage{url}

\usepackage{siunitx}     
\usepackage{amsmath}
\usepackage{tabularx}
\usepackage[table,xcdraw]{xcolor}
\usepackage{makecell}    
\usepackage{braket}      
\usepackage{bbold}       
\usepackage{hyperref}
\usepackage{dsfont}
\hypersetup{
    colorlinks=true,
    linkcolor=blue,
    filecolor=magenta,      
    urlcolor=blue,
    citecolor=blue,
    pdftitle={Error-detected gates},
    pdfpagemode=FullScreen,
    }

\begin{document}
\newcommand{\Hc}{\hat{\mathcal{H}}_c}
\newcommand{\tswap}{t_{\text{SWAP}}}
\newcommand{\gbs}{g_{\text{BS}}}
\newcommand{\fsr}{\Delta_{\text{FSR}}}
\newcommand{\Fdmm}{\mathcal{F_{\text{DMM}}}}
\newcommand{\iFdmm}{\mathcal{\overline{F}_{\text{DMM}}}}
\newcommand{\specialcell}[2][c]{%
\begin{tabular}[#1]{@{}c@{}}#2\end{tabular}}

\sisetup{range-phrase=--}
\title{Robust Quantum Communication through Lossy Microwave Links}

\author{James D. Teoh}
\email{ teoh@quantumcircuits.com}
\thanks{\\Current address: {Quantum Circuits, Inc., New Haven, CT, USA}}
\affiliation{Departments of Applied Physics and Physics, Yale University, New Haven, CT, USA}
\affiliation{Yale Quantum Institute, Yale University, New Haven, CT, USA}

\author{Nathana\"el Cottet}%
\affiliation{Departments of Applied Physics and Physics, Yale University, New Haven, CT, USA}
\affiliation{Yale Quantum Institute, Yale University, New Haven, CT, USA}

\author{Patrick Winkel}%
\affiliation{Departments of Applied Physics and Physics, Yale University, New Haven, CT, USA}
\affiliation{Yale Quantum Institute, Yale University, New Haven, CT, USA}

\author{Luke D. Burkhart}%
\affiliation{Departments of Applied Physics and Physics, Yale University, New Haven, CT, USA}
\affiliation{Yale Quantum Institute, Yale University, New Haven, CT, USA}

\author{Luigi Frunzio}%
\affiliation{Departments of Applied Physics and Physics, Yale University, New Haven, CT, USA}
\affiliation{Yale Quantum Institute, Yale University, New Haven, CT, USA}

\author{Robert J. Schoelkopf}%
\affiliation{Departments of Applied Physics and Physics, Yale University, New Haven, CT, USA}
\affiliation{Yale Quantum Institute, Yale University, New Haven, CT, USA}

\date{\today}

\begin{abstract}
Entanglement generation lies at the heart of many quantum networking protocols as it enables distributed and modular quantum computing. 
For superconducting qubits, entanglement fidelity is typically limited by photon loss in the links that connect these qubits together. 
We propose and realize a new scheme for heralded entanglement generation that almost entirely circumvents this limit. We produce Bell states with $92\pm1\%$ state fidelity, including state preparation and measurement (SPAM) errors, between separated superconducting bosonic qubits in a high-loss regime where direct deterministic state transfer fails. Our scheme exploits simple but fundamental physics found in microwave links, specifically the ability to treat our communication channel as a single standing wave mode. Combining this with local measurements on bosonically encoded qubits allows us to herald entanglement with success probabilities approaching the scheme's upper limit of 50\% per attempt.  
We then use the heralded Bell state as a resource to deterministically teleport a qubit between modules with an average state transfer fidelity of $90\pm1\%$. This is achieved despite the link possessing a direct single photon transfer efficiency of 2\%. 
Our work informs the design of future superconducting quantum networks, by demonstrating fast coupling rates and low loss links are no longer strict requirements for high-fidelity quantum communication in the microwave regime.
\end{abstract}
\maketitle

\section{Introduction}
\label{sec:introduction}\
In an ideal quantum network, communication between qubit modules neither limits the fidelity nor rate of quantum operations. Yet in reality, loss in the communication channel is the limiting factor for many quantum networking schemes.
 Optical fiber and free space channels are routinely used in a variety of platforms\,\cite{Northup2014,TeleportationReview2015,TeleportationReviewHu2023} but fiber coupling and photon collection inefficiencies have led to the development of probabilistic, heralded entanglement generation as the main communication resource\,\cite{QuantumTeleportation1993, QuantumRepeater1998}. 
In these approaches, high-fidelity entanglement and thus quantum communication is still possible, albeit at the cost of a reduced communication rate, often requiring hundreds or thousands of attempts until a single entangled state is heralded.

\indent In contrast, communication channels in the microwave regime are well-suited to superconducting qubit platforms and are typically either waveguides\,\cite{Wallraff5mFridge2020, Wallraff30mFridge2023, Grebel2024, Parth2024, Almanakly2025} or transmission lines\,\cite{Majer2007,2019ClelandRemote}. 
By engineering parametric\,\cite{Luke2020,2019ClelandRemote, 2019SchusterDarkMode, Zhou2024, Zhou2025} or resonant\,\cite{Majer2007} couplings, coupling inefficiencies can be negligible, enabling on-demand, deterministic entanglement generation and quantum state transfer. As noted by several authors\,\cite{Kurpiers2017,Wallraff5mFridge2020,Luke2020,Zhong2023, Qiu2025}, the loss in these links is comparable or even less than that of optical fibers, with attenuation lengths on the scale of several kilometers. Nevertheless, link loss is often still the dominant error source as the communication time is usually limited by the coupling rate to the link, which is often much slower than the photon `time-of-flight' through the link.
This error mechanism can be further exacerbated if lossy directional elements such as circulators are added to the communication channel as required by certain schemes\,\cite{Roch2014,Narla2016,AxlinePnC,PhillipePnC,Kurpiers2018PnC,Wallraff30mFridge2023,Wallraff5mFridge2020,2019WallraffTimeBin,irfan2025autonomousstabilizationremoteentanglement}.

\begin{figure*}[t] 
\includegraphics[width = 0.9\textwidth]{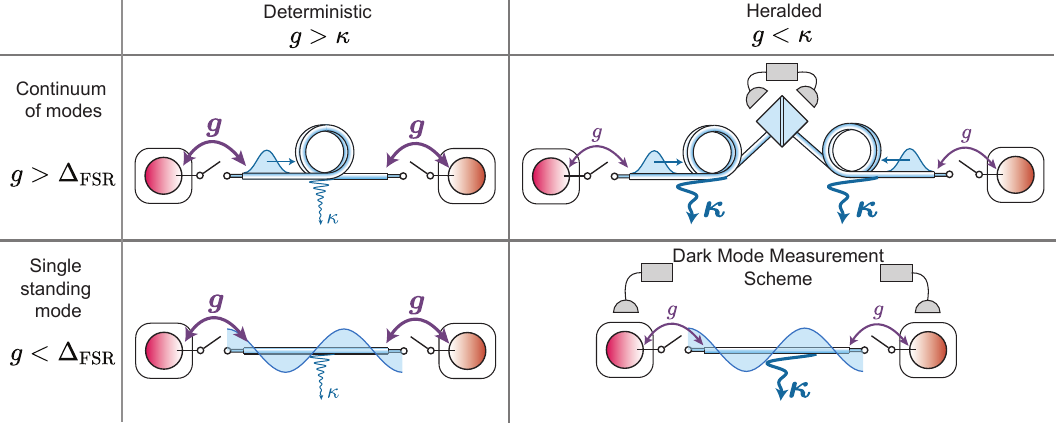}
\caption{\textbf{Approaches to quantum networking between two separated qubit modules.} The ratios of three physical rates determine the physics of our system and the suitability of different communication schemes. These are $g$, the coupling rate to the link, $\kappa$, the intrinsic link loss rate and $\fsr$, the free spectral range which is the frequency spacing between standing modes in the link.
For optical frequencies, we are restricted to the regime where  $g>\fsr$ (top row) and as such information must travel through the links in the form of wavepackets. With the addition of directional elements such as circulators, microwave links can also reach this regime. Over lab-scale distances, microwave links are typically in the regime where $g<\fsr$, allowing coupling to a single standing wave mode which may be used as a communication bus.   
When $g >\ \kappa$ (left column), deterministic communication is possible by directly transferring the qubit directly down the link as a flying wavepacket, such as in the CZKM scheme \cite{CZKM} (top left) or transferring via a bus mode (bottom left) when $g<\fsr$. When $g < \kappa$, we must design our schemes to be insensitive to link loss to achieve high fidelity communication, either by heralding (right column) or with adiabatic schemes. In our work (bottom right), we herald entanglement with local measurements in the regime $g<\kappa\ll\fsr$ in a novel scheme that exploits interference in the link mode to circumvent loss.}
\label{fig:Fig1}
\end{figure*}

\indent We focus on the task of overcoming these losses when networking superconducting qubit modules together over distances of 0.1--\SI{10}{\meter}, a scale useful for distributed quantum devices  within a single cryostat~\cite{Kevin2018TeleportedGate,LiangModularHelpsCosmicRays,laracuente2023modeling}. A distributed approach allows modules containing fewer  qubits to be independently fabricated and tested before joining them together. Over these distances, and for microwave wavelengths, it is easy to engineer a coupling to just one of the standing wave modes in the communication channel, which we call the `bus mode'. 
Compared to a continuum of modes\,\cite{CZKM}, the physics here is fundamentally different, and enables novel communication schemes that exploit the standing wave nature of this mode. These paradigms are compared in Fig.\,\ref{fig:Fig1}. 

\indent Previous works have used the discrete modes of a microwave link as a quantum bus\,\cite{Majer2007,Luke2020,2019ClelandRemote,2019SchusterDarkMode,2022ClelandEntanglementPurification,Zhong2023,Mollenhauer2025} but energy loss in the bus is typically what limits communication fidelity. 
Adiabatic schemes based on virtual Raman transitions\,\cite{Luke2020} or Stimulated Raman Adiabatic Passage (STIRAP)\,\cite{Cleland2020STIRAP} can in principle mitigate this loss but incur the penalty of an inherent slowdown in communication time which then increases the local qubit errors in the modules. Other works propose~\cite{AashClerk2018dissipative,Nori2021,Zapletal2022}  or demonstrate~\cite{irfan2025autonomousstabilizationremoteentanglement} that loss in the link can be used to engineer non-local dissipation between qubits to stabilize an  entangled state, but also incur a similar slowdown---in this case governed by the engineered dissipation rate.

\indent In this work, we present and realize a new scheme for heralded entanglement generation that mitigates the effects of a lossy bus mode, which we name the `Dark Mode Measurement' (\textbf{DMM}) scheme. To a very good approximation (see Appendix \ref{app:secret_loss}), bus loss is completely negligible on successful attempts and we do not suffer from an adiabatic slowdown that would otherwise make us susceptible to local qubit errors. Unlike optical heralding schemes, the success probability is independent of the channel loss rate and is inherently close to 50\% per attempt. 

\indent We use the DMM scheme to prepare high-fidelity  Bell states in the (two-legged) cat code, between superconducting cavity qubits housed in separate modules, and with a superconducting coaxial cable of length $\approx$ \SI{7}{\centi\meter} functioning as our communication channel. As shown in Fig.\,\ref{fig:Fig1}, our scheme combines the benefits of heralding to overcome loss with the simplicity of using a single standing wave mode as our communication channel. We measure a Bell state fidelity of $92\pm1\%$ with a success probability of $34.9\%$ per attempt and further make use of this state to deterministically teleport a state from one module to another with an average state transfer fidelity of $90\pm1\%$, despite our link having a single photon transfer efficiency of just 2\% (See Appendix~\ref{app:transfer_efficiency}). One may expect that engineering a large coupling rate to the link and/or minimizing link loss are essential to obtaining high communication fidelities. The DMM scheme demonstrates this is not strictly necessary, and offers a route to easing engineering requirements in future superconducting quantum networks.

\section{The Dark Mode Measurement Scheme}
\label{sec:scheme}
\subsection{Classical interference in a bus mode}
\indent In order to treat our microwave link as a single standing wave mode as opposed to a continuum of modes, we must couple our qubits to the link with characteristic coupling rate, $g$ such that $g<\fsr$, where $\fsr$ is the Free Spectral Range (FSR) defined as $\fsr=c/2L$,  where $c$ is the speed of light in vacuum and $L$ is the length of the microwave link connecting the two qubits. We label this mode as the `bus mode' which has energy loss rate, $\kappa_b$. Our scheme then utilizes interference effects between the two bosonic modes we wish to entangle, $\hat{a}_1$ and $\hat{a}_2$, which arise when we  simultaneously couple both modes to the bus mode, $\hat{b}$.
The desired couplings are simple beamsplitter interactions of the form
\begin{equation}
\label{eq:bs}
    \mathcal{\hat{H}}_{\text{BS},i}/\hbar = g_i\hat{a}_i\hat{b}^\dagger +g_i^*\hat{a}_i^\dagger\hat{b},
\end{equation}
for bosonic modes $i=1,2$ which reside in modules 1 and 2 respectively. In our experimental demonstration, these are the $\lambda/4$ fundamental cavity modes of two 3D microwave stub cavities.  By setting  $g_1=g_2=\gbs$ and actuating both couplings simultaneously we engineer the coupling Hamiltonian
\begin{equation}
\label{eq:hc}
    \Hc/\hbar=\gbs (\hat{a}_1+\hat{a}_2 ) \hat{b}^\dagger+h.c. 
\end{equation}
 Before describing the full scheme, we first examine the system dynamics for the case where the cavity modes are first prepared in a so-called `bright state' - two coherent states with the same amplitude and same relative phase. In the case where $\kappa_b=0$, if we enact $\Hc$ for time $\tswap=\pi/(2\sqrt{2}\,\gbs)$ (note the factor $\sqrt{2}$ speed enhancement, see App. \ref{app:Hct}) then all excitations in both cavity modes are fully transferred into the bus mode as follows:

\begin{equation}
    \ket{\alpha}_1\otimes\ket{0}_b\otimes\ket{\alpha}_2\rightarrow\ket{0}_1\otimes\ket{\sqrt{2}\alpha}_b\otimes\ket{0}_2,
\end{equation}
where the subscripts indicate the tensor product of states in cavity 1, bus, and cavity 2, respectively. These coherent states can be said to interfere constructively, a phenomenon that occurs whenever the coherent states have the same amplitude and same relative phase.

In contrast, we can prepare the system in a dark state which is completely oblivious to  dynamics associated with $\Hc$ and hence to any dissipation in the bus mode.
In general, a dark state of our system is any excitation of the dark mode $\hat{a}_d$, defined as 
\begin{equation}
\label{eq:dark_mode}
    \hat{a}_d = \frac{1}{\sqrt{2}}(\hat{a}_1-\hat{a}_2).
\end{equation}
The dark states of interest to us are coherent states with the same amplitude but \textit{opposite} relative phases such as the state $\ket{\alpha}_1\otimes\ket{0}_b\otimes\ket{-\alpha}_2$ which remain static under $\Hc$. As a result, no excitations are transferred into the bus mode after applying $\Hc$ for duration $\tswap$. \\
\indent For sufficiently large $|\alpha|$ we can then distinguish whether the cavities were initialized in a bright or dark state by performing local cavity measurements to determine whether or not the cavities are in vacuum. We refer to this measurement as the `vacuum check'. This is the measurement that gives the DMM scheme its namesake, where upon measuring that there are still excitations in the system, we also know the system must exist in a dark mode.
\subsection{Generating entanglement}
The strategy behind generating entanglement with the DMM scheme is to begin in a superposition of both bright and dark states which is initially separable and has no entanglement, but to carefully choose our starting states such that the dark states by themselves would be a maximally entangled state. Then, at the end of the protocol, the vacuum check measurements can be used to herald an entangled state whenever we measure the system to be in a dark state. Since this state never interacted with the bus mode, it is completely unaffected by loss in the bus mode. 

\indent In its simplest version, we achieve this by first initializing two local (two-legged) cat states in each cavity in an even superposition in the state 
\begin{equation}
    (\ket{\alpha}+\ket{-\alpha})_1\otimes\ket{0}_b\otimes(\ket{\alpha}+\ket{-\alpha})_2,
\end{equation}
where we have omitted the normalization factors for brevity. Dropping the tensor products from our ket notation as well, we can rewrite this state as
\begin{equation}
    \ket{\alpha,0,\alpha}+\ket{-\alpha,0,-\alpha}+\ket{\alpha,0,-\alpha}+\ket{-\alpha,0,\alpha},
\end{equation}
where we can clearly see this state consists of a superposition of two bright coherent states and two dark coherent states.
After enacting $\Hc$ for time $t_{\text{SWAP}}$, the state evolves to
\begin{align}
    \ket{0,\sqrt{2}\alpha,0}+\ket{0,-\sqrt{2}\alpha,0}+\ket{\alpha,0,-\alpha}+\ket{-\alpha,0,\alpha}.
\end{align}
Finally we perform the vacuum check. If the measurement outcome reveals both cavities are \textit{not} in the vacuum state, the state is projected into an entangled state. In the limit of large $|\alpha|$, where the overlap with the vacuum state is negligible (See Appendix \ref{app:finite_alpha}), this state can be written as
\begin{align}
\ket{\alpha,0,-\alpha}+\ket{-\alpha,0,\alpha}.
\end{align}
 This state is a Bell state in the two-legged cat code, also referred to as a `cat-in-two-boxes' state, and has been previously studied\,\cite{ChenC2B,chapman2022high} with `local' superconducting qubit hardware, where both cavities reside in the same module. We expect to obtain this state with 50\% success probability per attempt. When we obtain the other outcome, we know the cavities are both empty and have transferred all their energy to the lossy bus mode. We must reset the bus mode to vacuum, by waiting, reset the cavities, and retry the scheme until entanglement is successfully heralded. 

\indent The remarkable feature of this scheme is that it proceeds in mostly the same way even when there is significant energy loss in the bus mode. Regardless of the value of $\kappa_b$, the dark modes are always static under $\Hc$ and hence loss in the bus does not degrade entanglement fidelity. If anything, a larger $\kappa_b$ may help to quickly reset the bus mode after failed attempts, which is necessary before the next attempt. However, $\tswap$, the time taken for bright states to fully transfer into the bus mode may become prohibitively long if we are in the regime where $\kappa_b\gg\gbs$. These different loss regimes are explored further in simulations presented in App. \ref{app:Hct}.
Experimentally, we also work with amplitudes of $\alpha$ where the finite overlap with vacuum, $\braket{0|\alpha}$ cannot be ignored. However, as described in App.\,\ref{app:finite_alpha}, the scheme can be easily adjusted to still produce perfect Bell states, provided we carefully redefine the basis of our logical codewords. This finite overlap also leads to an additional decrease in success probability with decreasing $\alpha$, since the dark states now have a non-negligible overlap with the vacuum state. 
\newpage
\section{Experimental Implementation}
\begin{figure}[] 
\includegraphics[width = 1.0\columnwidth]{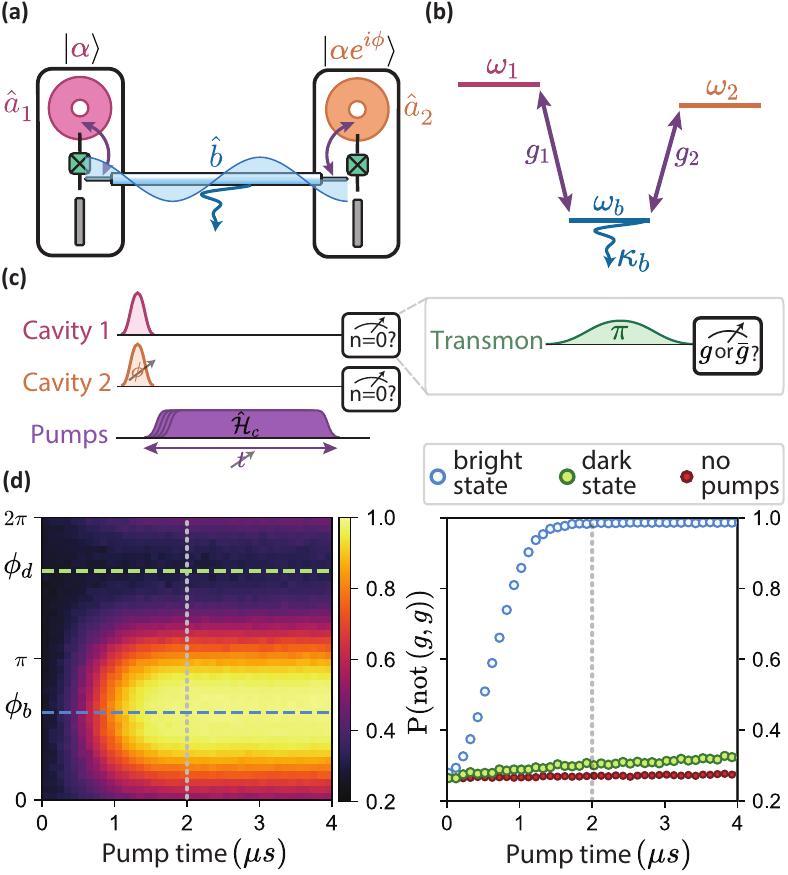}
\caption{\textbf{Measuring dynamics of bright and dark states} 
\textbf{a)} Superconducting hardware implementation used for all experiments. Each module contains a 3D stub cavity (colored pink or orange), with cavity mode $\hat{a}_i$, a transmon (green) and a readout resonator (grey). Parametric conversion between each cavity mode and the lossy bus mode (blue), $\hat{b}$, is actuated via a four-wave mixing processes, by applying two microwave pump tones through the transmon port (not shown, see App. \ref{app:wiring}). Transmons also enable state preparation, QND measurements and tomography. 
\textbf{b)} Energy level diagram and couplings. We tune $g_1=g_2$ in-situ with drive amplitudes and activate these couplings simultaneously.
\textbf{c)} Pulse sequence for characterizing the DMM scheme with classical bright and dark states. Cavities are initially displaced to the coherent states $\ket{\alpha}\ket{\alpha e^{i\phi}}$ where the relative phase, $\phi$, is swept. We apply the coupling Hamiltonian, $\Hc$ for variable pump time, $t$, and then measure whether either cavity is in vacuum (the `vacuum check'). Local measurements are carried out via number selective $\pi$-pulses on each transmon, followed by standard dispersive readout of each transmon. Transmons are excited only if the cavity contains 0 photons, making this a QND measurement. We declare a dark state (pass) when we obtain the outcome $(g,g)$ and a bright state otherwise (fail). 
\textbf{d)} Dynamics of bright and dark states under $\Hc$. We measure P$(\text{not} (g,g)$), the probability we fail the vacuum check and identify phase $\phi = \phi_b$ as the phase of the bright state which decays to vacuum the fastest (bright regions). At $\phi=\phi_d = \phi_b \pm \pi$ dynamics are static (dark regions), indicating we have prepared the dark state. 
With line cuts, we contrast the dynamics of bright and dark states. In the absence of pumps (red circles), dark states have a finite probability (0.27) of failing the vacuum check, due to their overlap with vacuum. Transmon heating by the pumps steadily increases this probability in time (green circles).
 Most importantly, after pumping for  $>$\SI{2}{\micro\second}, bright states fail the vacuum check with high probability ($>0.98$).
 }

\label{fig:Fig2}
\end{figure}

\begin{figure*}[] 
\includegraphics[width=0.85 \linewidth]{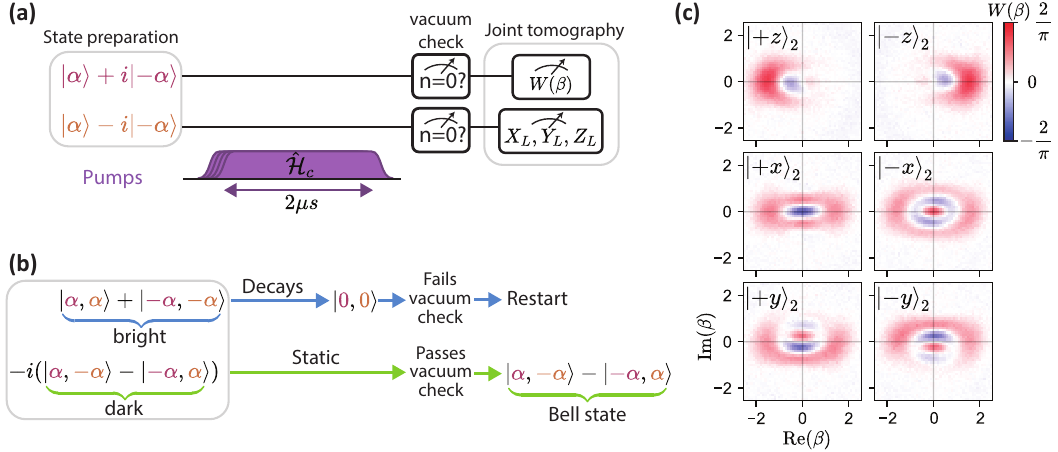}
\caption{\textbf{Dark Mode Measurement for entanglement generation} \textbf{a)} Experimental pulse sequence for entanglement generation and two-qubit tomography. The pulse sequence is identical to Fig.\,\ref{fig:Fig2}c but now starting from two local Schr\"odinger cat states. Cavity 2 is measured in its logical basis by mapping its state onto transmon 2 via optimal control pulses whilst simultaneously measuring the Wigner function of cavity 1. \textbf{b)} Differing evolution of bright and dark states under $\Hc$ allows the vacuum check to herald entanglement. The bright state quickly decays to vacuum under $\Hc$  and thus fails the vacuum check. The dark state, which is also a Bell state in the cat code, is static under $\Hc$ and passes the vacuum check, resulting in a heralded Bell state between the modules. \textbf{c)} Measured Wigner functions of cavity 1, for $\alpha=1.414$, conditioned on the outcome of cavity 2 measurements in each of the three logical bases (outcome indicated by the inset). Wigner functions show strong anticorrelation in all three bases, as expected for this particular Bell state. As a consequence of the vacuum check, the logical basis of cavity 1 is a modified version of the cat code as defined in App. \ref{app:finite_alpha}.
}
\label{fig:Fig3}
\end{figure*}

\indent We realize our scheme in a superconducting circuit platform where we generate entanglement between two spatially separated qubit modules. Each module contains a 3D high-Q microwave cavity mode in which a qubit is bosonically encoded in the cat code. Modules are joined together by a short length (\SI{6.7}{cm}) of commercially available Niobium Titanium (NbTi) superconducting coaxial cable, which serves as our communication channel. A transmon is contained within each module with a dispersive coupling to both the cavity and bus mode. 

\indent Importantly, each transmon provides the requisite non-linearity for multiple purposes. Firstly, we use it as an ancilla in the  preparation of cat states in each cavity. Secondly, by applying two off-resonant microwave pump tones we can use it as a mixer to actuate tunable beamsplitter couplings between the cavity and the bus mode via a four-wave mixing process\,\cite{YaxingBilinear,YvonneProgrammable2018,Luke2020} to engineer the Hamiltonian $\Hc$. Thirdly, number selective $\pi$-pulses applied to the transmon~\cite{LeghtasqcMAP}, followed by transmon readout, suffice as our Quantum Non-Demolition (QND) `vacuum check' measurement and finally, the transmons can be used to perform two-qubit quantum state tomography to quantify the  entanglement between the cavity modes.

We choose to operate close to the `critical damping' regime of the DMM scheme to minimize the time required to generate entanglement and to allow fast bus reset between attempts. (See App. \ref{app:DMM_details} and \ref{app:mr} for more details). We set $g_1/2\pi=g_2/2\pi=160\pm5\,\text{kHz}$ by tuning our microwave pumps in-situ via a procedure described in App. \ref{app:pump_cal} and engineer the loss in our bus mode to be $\kappa_b/2\pi=600\pm10\,\text{kHz}$ through use of a 3D Purcell filter (See App. \ref{app:Purcell}).  The cavities, which we refer to as cavity 1 and cavity 2 are etched 99.999\% purity Aluminum stub cavities with $T_1$ times of \SI{385}{\micro\second}  and \SI{520}{\micro\second} respectively, two orders of magnitude longer than the duration of our DMM scheme. Their mode frequencies are \SI{6509}{MHz} and \SI{6487}{MHz} respectively. We use the $3\lambda/2$ harmonic of the cable as our bus mode, with a frequency of \SI{5658}{MHz}, a $T_1$ of \SI{270}{\nano\second} and $\fsr\sim$~\SI{2}{GHz}. A full list of system parameters can be found in App. \ref{app:params}.
A detuning $\sim$~\SI{1}{GHz} between the cavities and bus mode avoids spurious couplings and ensures a high on-off ratio. In the absence of the coupling pumps there is no measurable exchange of energy between the cavities and the bus mode and there is no indication that cavity lifetimes are Purcell-limited by the lossy bus mode. The hardware and mode couplings are illustrated in Fig.\,\ref{fig:Fig2}a and \ref{fig:Fig2}b respectively.

Capacitive coupling between the transmon and the cable is necessary to actuate $\Hc$ via four-wave mixing, and is achieved by stripping back the outer conductor and Teflon dielectric to expose the inner conductor (\SI{8}{mm} on each side), which protrudes above one of the transmon's antenna pads. No wire bonding is required. We clamp the cable to each module as described in\,\cite{Luke2020}, allowing the cable to be easily adjusted or exchanged between fridge cooldowns. 

\indent We first characterize the performance of our scheme with coherent states as shown in Fig.\,\ref{fig:Fig2}, starting with states of the form $\ket{\alpha}\ket{\alpha e^{i\phi}}$ in both cavities, as shown in Fig.\,\ref{fig:Fig2}, where $\phi$ is the relative phase between the two cavity displacements. We set $\alpha = 1.414\approx\sqrt{2}$, the value at which we measure the highest entanglement fidelity. Whilst we cannot produce entanglement when starting from these classical coherent states, we can quickly verify the expected dynamics of the bright and dark states under $\Hc$ and dissipation in the bus mode, wherein dark states are unaffected but bright states decay to vacuum via the lossy bus mode in around \SI{2}{\micro\second}.

\indent We `pass' the vacuum check if both transmons are measured in their ground states, $(g,g)$ after the selective $\pi$-pulses are applied. This should indicate the cavities are in a dark state. The probability we achieve this result given we prepared either a bright or dark state essentially quantifies how well the DMM scheme can work in the absence of SPAM errors or local errors such as cavity photon loss. (See App. \ref{app:budget}.) We quantify this with a number we call the dark mode measurement infidelity, $\iFdmm$, which we define as  a `false positive' rate i.e. the chance we erroneously accept a bright state when we pass the vacuum check. This probability is given by Bayes' rule as
\begin{equation}
    \iFdmm= \frac{p(gg|\text{bright})}{p(gg|\text{dark})+p(gg|\text{bright})},
\end{equation}
where $p(gg|\text{bright})$ is the probability we pass the vacuum check given we begin in a bright state, and $p(gg|\text{dark})$ is the probability we pass the vacuum check given we began in a dark state.
\indent We measure $\iFdmm = 2.1\%$, a value larger than what should be expected from transmon decoherence during the selective $\pi$-pulses alone (see App. \ref{app:budget}), but still small enough to not be the dominant error source in our experimental implementation.

\section{Entanglement Generation}

To generate entanglement we use the exact same pulse sequence as in Fig.\,\ref{fig:Fig2}c, except we now initially prepare the two un-entangled cat states in each cavity, which can be written as
\begin{equation}
\ket{\psi_{\text{init}}}\propto(\ket{\alpha}+i\ket{-\alpha})_1\otimes\ket{0}_b\otimes(\ket{\alpha}-i\ket{-\alpha})_2    
\end{equation}
We chose these particular so-called `parity-less' cat states for ease of state preparation as further discussed in App. \ref{app:cat_prep}. These cat states are prepared by first displacing the cavities to the state $\ket{\alpha}\ket{-\alpha}$, followed by pulses applied to each transmon that evolve the cavity states to $\ket{\alpha}\rightarrow\ket{\alpha}+i\ket{-\alpha}$ and $\ket{-\alpha}\rightarrow\ket{\alpha}-i\ket{-\alpha}$ (normalization factors omitted) whilst also returning the transmon to its ground state. These pulses are found through optimal control theory (OCT) pulses found via the gradient ascent algorithm \cite{Heeres2017} and can be viewed as numerically optimized Selective Number Arbitrary Phase (SNAP) pulses\,\cite{SNAP_Heeres_2015}. As such, we refer to them as OCT SNAP pulses.  

After state preparation, we actuate $\Hc$ by applying the pump tones for \SI{2}{\micro\second}. Successful entanglement is heralded when we obtain the outcome $(g,g)$ and pass the vacuum check, which ideally leaves us in the pure entangled state
\begin{equation}
\label{eq:final_state}
\ket{\psi_{\text{ent}}}\propto\hat{\Pi}_{\bar{0}\bar{0}}(\ket{\;\alpha\;,\;0\;,-\alpha}-\ket{-\alpha,\;0\;,\;\alpha\;}),    
\end{equation}
where $\hat{\Pi}_{\bar{0}\bar{0}}$ is the `not-vacuum' projector defined as
\begin{equation}
\hat{\Pi}_{\bar{0}\bar{0}}=(\mathds{1}-\ket{0}\bra{0})_1\otimes \mathds{1}_b\otimes (\mathds{1}-\ket{0}\bra{0})_2, 
\end{equation}
which results from our vacuum check `eliminating vacuum' from the dark state, requiring us to redefine our logical code words as
\begin{align}
    \begin{split}
        \ket{+}_L &\propto (\mathds{1}-\ket{0}\bra{0})\left(\ket{\alpha} + \ket{-\alpha}\right)\\
        \ket{-}_L &\propto \ket{\alpha} - \ket{-\alpha}.
    \end{split}
\end{align}
Crucially, the state in Eq. \ref{eq:final_state} remains a perfect Bell state in this modified basis. The finite overlap between our dark state and vacuum means our success probability is now less than one half and is instead given by
\begin{equation}    p_{\text{success}}=\frac{1}{2}\left(1-2e^{-|\alpha|^2}+e^{-2|\alpha|^2}\right),
\end{equation}
which places an upper bound of $\approx37.4\%$ on the success probability per attempt at $\alpha = \sqrt{2}$. If we obtain any measurement outcome other than $(g,g)$, we have `failed' the vacuum check. We reset the system and proceed with the next experimental shot. If we obtain the $(g,g)$ outcome we have `passed' the vacuum check and immediately proceed with two-qubit tomography. 
\begin{figure}[] 
\includegraphics[width=0.7 \linewidth]{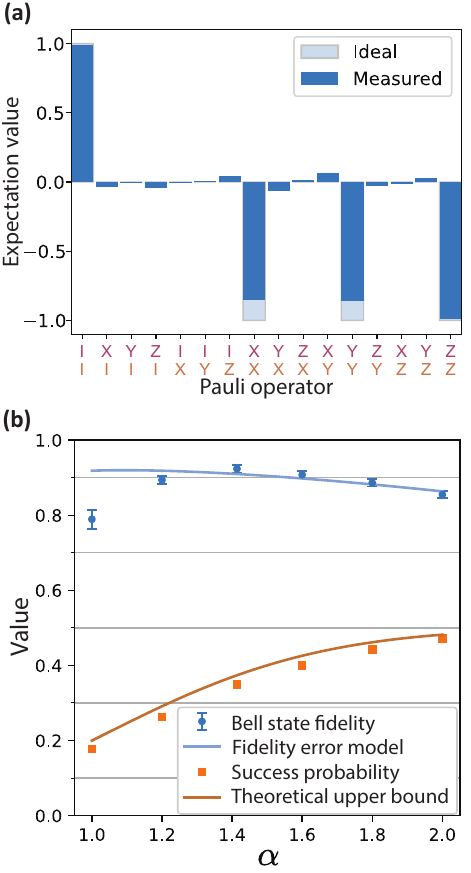}
\caption{\textbf{Measured Joint Pauli correlations and Bell state fidelities for different coherent state amplitudes.} \textbf{a)} Expectation values of joint Pauli correlations, extracted from the Wigner function data presented in Fig.\,\ref{fig:Fig3}c at $\alpha = 1.414$, for which the highest Bell state fidelity was measured. For our ideal Bell state, the joint Pauli correlations $\braket{II}$, $\braket{XX}$, $\braket{YY}$ and $\braket{ZZ}$ have a magnitude of one and are the only nonzero correlations. Smaller expectation values measured for the $\braket{XX}$ and $\braket{YY}$ Pauli operators indicate photon loss in the cavity modes is the dominant error source. \textbf{b)} Measured Bell state fidelities and scheme success probabilities for varying values of $\alpha$. Solid blue line indicates the expected Bell state fidelity due to our error model, which includes cavity decay, vacuum check fidelity and the decode pulse. (See App. \ref{app:budget}.) The solid orange line is the success probability from Eq. \ref{eq:true_success_prob}, which accounts for the finite $\braket{\alpha|0}$ overlap.   
}
\label{fig:Fig4}
\end{figure}
\\
\indent We reconstruct the state in its logical basis and quantify the entanglement fidelity following the approach developed in \cite{Luke2020} and outlined in App. \ref{app:twobit_tomo}, which avoids having to measure the full 4D joint-Wigner function\,\cite{ChenC2B}. We measure cavity 2 in one of its three logical bases $\{X_L,Y_L,Z_L\}$ by first mapping the logical qubit state from cavity 2 onto the $g$-$e$ manifold of transmon 2. This is achieved via an OCT pulse which we call the decode pulse. This additional step slightly degrades our measured entanglement fidelity but greatly simplifies our two-qubit tomography and analysis. By applying the appropriate transmon 2 rotations via Rabi drives we can then read out transmon 2 in all three measurement bases. Rather than decoding cavity 1 onto its transmon as well (which would further increase SPAM error), we instead simultaneously measure its Wigner function. As shown by Fig.\,\ref{fig:Fig3}c,  filtering the Wigner function conditioned on the different cavity 2 logical measurement outcomes visually reveals the correlations we would expected for an entangled Bell state between the cavity modes in the modified cat basis. This also allows us to extract the expectation values of the joint Pauli operators as shown in Fig.\,\ref{fig:Fig4}a.  From these values, we can extract the logical two-qubit density matrix, and finally the Bell state fidelity, $\mathcal{F}_{\text{Bell}}$. Performing this logical state tomography also reveals the dominant error source is likely cavity photon loss, which agrees well with our error budget. (See App. \ref{app:budget})

We determine $\mathcal{F}_{\text{Bell}}$ at various values of $\alpha$, as shown by Fig.\,\ref{fig:Fig4}b (see App. {\ref{app:ent_ext}} for full data), and find the highest Bell state fidelity is obtained for $\alpha = 1.414\approx\sqrt{2}$, with fidelity $\mathcal{F}_{\text{Bell}}=92\pm1\%$, including SPAM errors. At larger values of $\alpha$, cavity photon loss becomes more likely. At smaller values, the vacuum check reliability decreases (along with the success probability) and we observe more unwanted leakage to the vacuum state. These two effects result in an optimum value of $\alpha$ for which Bell state fidelity is maximized. The success probability per attempt also varies predictably with $\alpha$, approaching its theoretical maximum value of $50\%$ as $\alpha$ increases.
\section{Quantum State Teleportation}
\begin{figure}[] 
\includegraphics[width=1.0 \linewidth]{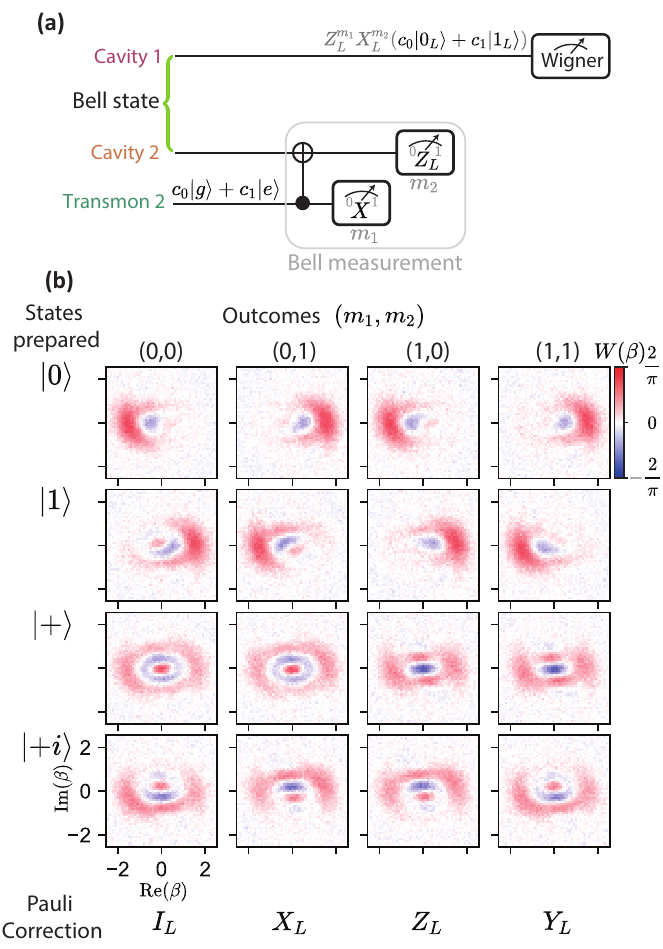}
\caption{\textbf{Deterministic Teleportation through a very lossy link} \textbf{a)} Implementation of the quantum teleportation circuit. Immediately after we herald a Bell state from our vacuum checks, we prepare the state we wish to teleport in transmon 2. We then perform a Bell measurement in module 2 to obtain two outcomes $m_1$ and $m_2$ that inform us of the Pauli corrections/updates that would need to be applied to cavity 1. (See App. \ref{app:Bell_msmts} for exact implementation.) \textbf{b)} Measured Wigner functions for cavity 1 showing quantum state transfer via teleportation. Columns show the measured Wigner function for cavity 1, filtered by the four different possible Bell measurement outcomes, $(m_1,m_2)$, each equally likely. Each row shows a different cardinal state (initially prepared in transmon 2) being teleported to cavity 1. We actively correct for Kerr distortions (see App. \ref{app:Kerr_comp}) but some residual distortion is still visually noticeable in the Wigner functions.
}
\label{fig:Fig5}
\end{figure}

Our high bus loss precludes the use of any scheme for direct quantum state transfer. However, we can instead  use our high fidelity Bell state as a resource for quantum state transfer via teleportation~\cite{QuantumRepeater1998}, which circumvents the limits imposed by our bus loss.  We implement the `textbook' scheme for deterministic quantum state teleportation\,\cite{QuantumTeleportation1993} between our qubit modules, as shown in Fig.\,\ref{fig:Fig5}a.

To perform quantum teleportation, we need a logical qubit state to teleport. Transmon 2 serves as our initial qubit, which is teleported to cavity 1 at the end of the protocol. Once we have successfully heralded a Bell state, we prepare the logical qubit state in the $g$-$e$ manifold of transmon 2. This state can be written as
\begin{equation}
    c_0\ket{g}+c_1\ket{e}.
\end{equation}
We teleport this qubit state to cavity 1 by performing a Bell measurement in module 2 between cavity 2 and transmon 2, which requires a local entangling gate between these modes (see App. \ref{app:Bell_msmts}.) After this measurement the qubit is now encoded in the modified cat basis in cavity 1 as
\begin{equation}
   c_0\ket{0_L}+c_1\ket{1_L} 
\end{equation}
 up to known Pauli corrections that are determined from the outcomes of the Bell measurement. These corrections may be applied with subsequent gates on the cavities (active teleportation) or tracked in software (passive teleportation)\,\cite{TeleportationReview2015}. We opt for the latter and as such we can see the Pauli corrections needed to recover cavity 1 by measuring its Wigner function conditioned on the four Bell measurement outcomes as shown in Fig.\,\ref{fig:Fig5}b.

Our teleportation is also deterministic, only the heralding of the Bell state is probabilistic---once we have a Bell state we keep 100\% of the data when we perform teleportation. We find an average state transfer fidelity of $\mathcal{F}_{\text{QST}}=90\pm1\%$, averaging over all Bell measurement outcomes (see App.\,\ref{app:tele_processing}) and the four cardinal qubit states we attempt to transfer, which are $\{\ket{0},\ket{1}, \ket{+},\ket{+i}\}$. When we initially prepare the qubit in transmon 2, we choose $\ket{0} =\ket{g}$ and $\ket{1} =\ket{e}$. The new basis that encodes the qubit once it has been teleported to cavity 1 is defined in App.\,\ref{app:DMM_details}. When we compare $\mathcal{F}_{\text{QST}}$ to $\mathcal{F}_{\text{Bell}}$ we infer that the extra steps required to perform the Bell measurement introduce an additional $\sim 2\%$ error.

\section{Discussion and Outlook}
We have shown high-fidelity quantum communication and state transfer is possible in the most demanding experimental regime, where the loss rate of the bus far exceeds the rate at which we can transfer energy/information into (or out of) the bus. By combining the interference effects afforded by a standing wave communication channel with  heralding measurements, we have demonstrated high fidelity Bell state generation and have further used this to teleport arbitrary qubit states across spatially separated modules. We emphasize that unlike  optical heralding, our scheme's performance is largely agnostic of the loss rate of the communication channel, with the probability of success inherently close to 50\% per attempt. 
\indent Further improvements to the Bell state fidelity are expected to result from improving the coherence times of modes \textit{within} our quantum modules, since loss in the bus mode does not limit this scheme's performance. 

The DMM scheme shows, rather counterintuitively, that non-local but lossy interactions can indeed produce close-to-perfect entanglement, without resorting to adiabatic schemes and their associated slowdowns. Our work lends support to the idea that engineering ever-lower loss links and faster couplings is no longer the only way to boost communication fidelities, provided we are willing to forgo on-demand communication due to the heralding requirement. The DMM scheme in principle enables the use of a wider range of physical links which would otherwise be considered too lossy for direct microwave communication. Examples of novel links that are viable with this scheme include non-superconducting links made from materials such as copper, microwave switches, and printed-circuit boards with onboard routing for reconfigurable bus modes. Our scheme may also be useful for related qubit platforms such as hybrid superconducting systems with phononic modes, where it may be hard to engineer $g\gg\kappa_b$ due to losses in the phononic modes or difficulty engineering a strong coupling between qubits and the link.
\\

\section*{Acknowledgements}
We thank Neel Thakur for their detailed feedback on the manuscript. We thank Benjamin Chapman for providing the SNAIL parametric amplifiers used in the experiment.

This research was sponsored by the Army Research Office (ARO) under grant no. W911NF-23-1-0051, by the Air Force Office of Scientific Research (AFOSR) under award number FA9550-21-1-0209. The views and conclusions contained in this document are those of the authors and should not be interpreted as representing the official policies, either expressed or implied, of the ARO, AFOSR or the US Government. The US Government is authorized to reproduce and distribute reprints for Government purposes notwithstanding any copyright notation herein. Fabrication facilities use was supported by the Yale Institute for Nanoscience and Quantum Engineering (YINQE) and the Yale University Cleanroom.

L.F. and R.J.S. are founders and shareholders of Quantum Circuits Inc. (QCI).

\bibliography{main}

\begin{titlepage}\centering
\vspace*{\fill}
\LARGE \textbf{Supplementary Information}
\vspace*{\fill}
\end{titlepage}

\appendix
\onecolumngrid

\section{Details of the DMM scheme}
\label{app:DMM_details}
\subsection{Duration to enact $\Hc$ for different bus loss regimes}
\label{app:Hct}
\begin{figure*}[h] 
\includegraphics[scale = 0.7]{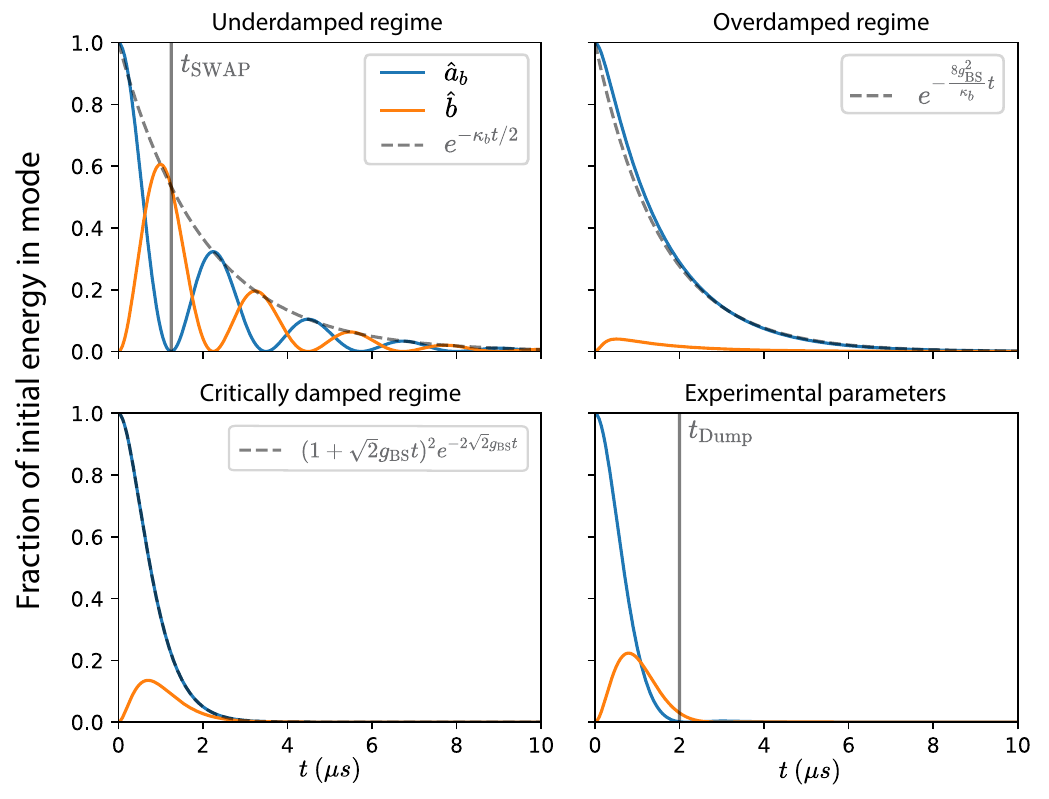}
\caption{\textbf{The DMM scheme in different bus loss regimes.} The ratio of $\kappa_b$ to $\gbs$ determines both how long we must apply $\Hc$ and also how long we must wait for the bus mode to reset between failed attempts. We assume the cavity states initially start entirely in a bright state and plot the energy decay dynamics for four regimes of interest. The dynamics are generated from Eq. \ref{eq:Heisenberg} with the initial conditions $a_d(0) = 0, a_b(0) = 1,b(0) = 0$. We fix $\gbs/2\pi=$\SI{160}{kHz} and vary $\kappa_b/2\pi$ to be \SI{160}{kHz} (underdamped), \SI{2000}{kHz} (overdamped), \SI{905}{kHz} (critically damped) or \SI{600}{kHz} (experimental parameters). 
Dashed lines are expressions for the energy decay rate of the bright mode, given by the analytical solutions to Eq. \ref{eq:Heisenberg}.
Critical damping exhibits the fastest energy decay from the system and is reached when $\kappa_b = 4\sqrt{2}\gbs$. For our experiment, $\kappa_b\approx4\gbs$ which is sufficiently close to the critically damped regime. Entanglement is generated the fastest when we are in the underdamped regime, as we are able to perform the vacuum check after time $t_{\text{SWAP}}$. In the experiment, we begin the vacuum check after applying $\Hc$ for time $t_{\text{Dump}}=$\SI{2}{\micro\second}. Any residual photons in the bus quickly decay during the \SI{1}{\micro\second} duration of the vacuum check and we do not measure any improvement if we increase this wait time.}
\label{fig:sfig_regimes}
\end{figure*}

\noindent In the case where $\kappa_b=0$, we must enact $\Hc$ for time 
\begin{equation}
 \tswap=\frac{\pi}{2\sqrt{2}\gbs}   
\end{equation}
This duration is a factor of $\sqrt{2}$ shorter than the time to swap two bosonic modes under the Hamiltonian in Eq.\ref{eq:bs}. The origin of this speedup becomes apparent by writing Eq. \ref{eq:hc} as
\begin{equation}
\label{eq:hc_r2}
    \Hc/\hbar=\sqrt{2}\gbs \frac{(\hat{a}_1+\hat{a}_2 )}{\sqrt{2}} \hat{b}^\dagger+h.c. 
\end{equation}
and defining the new mode basis 
\begin{align}
    \begin{split}
        \hat{a}_b &= \frac{(\hat{a}_1+\hat{a}_2 )}{\sqrt{2}}\\
        \hat{a}_d &= \frac{(\hat{a}_1-\hat{a}_2 )}{\sqrt{2}}\\
        \hat{b} &= \hat{b}
    \end{split}
\end{align}
We refer to $\hat{a}_b,\,\hat{a}_d,\,\hat{b}$ as the bright mode, dark mode and bus mode respectively. This basis does not represent the normal modes of $\Hc$ but is useful for analyzing system dynamics.  Bright states refer to states that only populate the bright mode and this is the only mode that couples to the bus mode. Similarly, dark states only populate the dark mode and have no coupling with the bus mode. We can rewrite Eq. \ref{eq:hc_r2} as
\begin{equation}
\label{eq:hc_bright}
    \Hc/\hbar=\sqrt{2}\gbs\:\hat{a}_b \hat{b}^\dagger+h.c. 
\end{equation}
to see how the $\sqrt{2}$ speedup appears. 
\\
\\
When $\kappa_b\neq0<\gbs$ we are in the under-damped regime, since many oscillations between modes $\hat{a}_b$ and $\hat{b}$ persist until all energy has decayed from these modes. This regime produces entanglement the fastest, by allowing us to perform the vacuum check immediately after applying $\Hc$ for time $\tswap$. 
\\
\\
In principle, we do not need to wait for the bus mode to empty before performing the vacuum check. However, for failed attempts, we must wait several decay constants for the bus mode to reset to vacuum, at decay rate $\kappa_b$ before trying again. For our specific experimental implementation, we must wait for the bus mode to be empty before performing the vacuum check, which motivates us to use the critical coupling regime.
\\
\\
We enter the critical coupling regime as $\kappa_b$ approaches $4\sqrt{2}\gbs$. Rather than lightly damped oscillations, both modes (the bright mode, $\hat{a}_b$ and the bus mode $\hat{b}$)  decay exponentially to vacuum at rate $\kappa_b/2$, and we must apply $\Hc$ for the duration of several decay constants until we can perform the vacuum check. This is the regime for which the experiment was performed, where we wait for approximately 4 decay constants.
\\
\\
When $\kappa_b\gg\gbs$, we enter the over-damped regime where now the loss in the bus affects the entanglement speed more severely. In this regime, both modes decay at reduced rate $2\gbs^2/\kappa_b$ requiring us to apply $\Hc$ for longer in order for the bright mode to reach vacuum. Mode dynamics, as well as the duration we must enact $\Hc$ are plotted in Fig.\,\ref{fig:sfig_regimes} below for these three regimes.
\\
\\
Dynamics of the field operators can be obtained by solving the following (classical) system of equations:
\begin{align}
\label{eq:Heisenberg}
\begin{split}
    \dot{a}_d(t) &= 0\\
    \dot{a}_b(t) & = i \sqrt{2}\, \gbs\:b(t)\\
    \dot{b}(t) & = -i \sqrt{2}\, \gbs\:a_b(t)-\frac{\kappa_b}{2}\, b(t)
\end{split}
\end{align}
\subsection{The DMM scheme at finite $\alpha$}
\label{app:finite_alpha}
\begin{figure*}[h] 
\includegraphics[scale = 0.45]{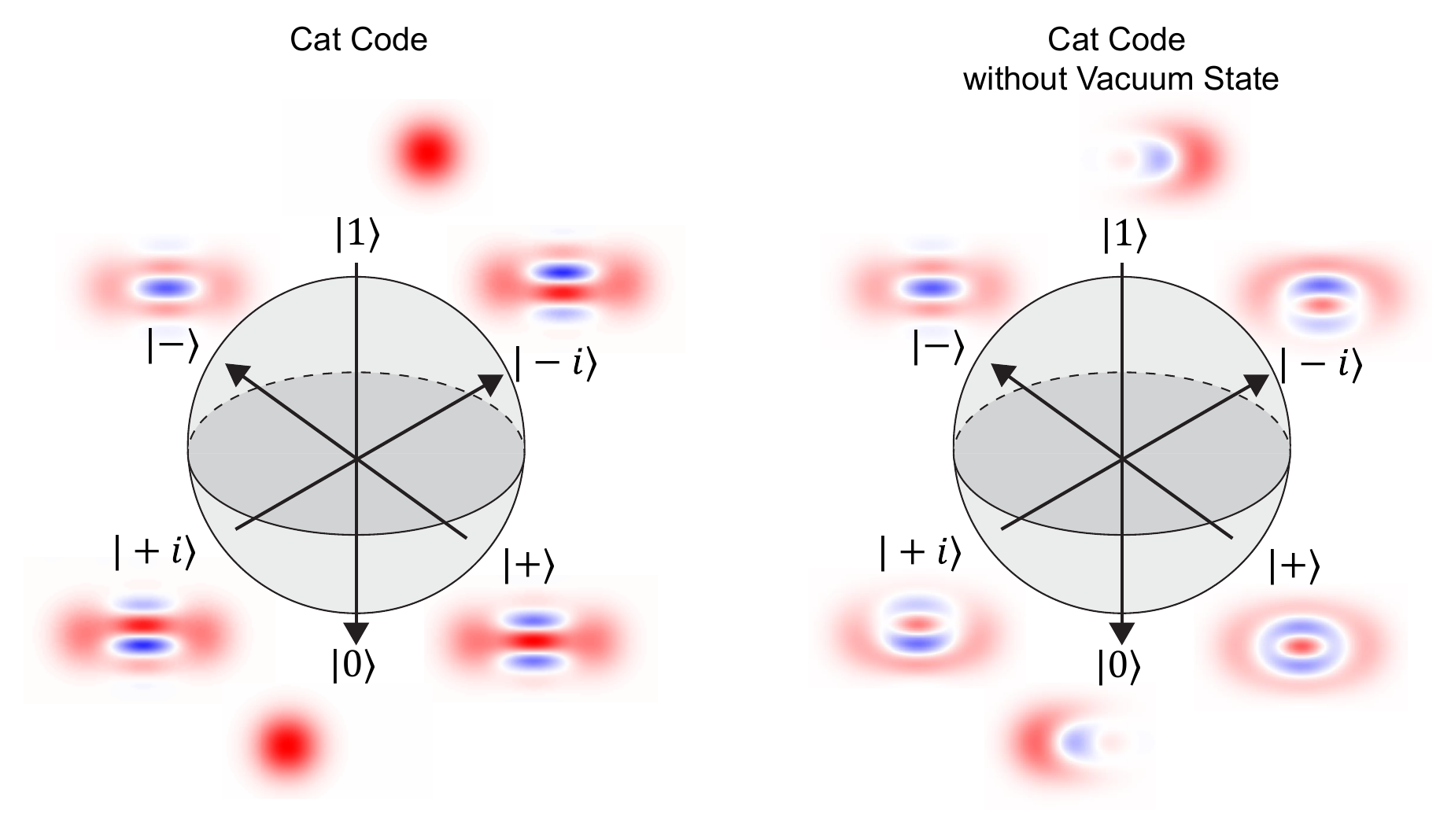}
\caption{\textbf{Modifying the codewords for small $\alpha$ after the vacuum check.} On the left we show the logical Bloch sphere for the standard 2-legged cat code and compare it to the Bloch sphere on the right which is for the modified basis and has the $\ket{0}$ state removed from the codewords. Here we show both encodings for $\alpha=\sqrt{2}$. The vacuum check has the effect of removing $\ket{0}$ from the dark state but by carefully redefining our logical basis, we can ensure our scheme still produces perfect Bell states. }
\label{fig:Spheres2}
\end{figure*}

\noindent How does working at small values of $\alpha$ affect the DMM scheme? We define small to be values of $\alpha$ where the overlap $|\braket{0|\alpha}|^2=e^{-|\alpha|^2}$ cannot be ignored, i.e. larger than errors due to decoherence. For $\alpha=2$, this overlap error is $\approx2\%$ and for our optimal experimental value of $\alpha=\sqrt{2}$ this is $\approx 13.5 \%$. There are two important consequences of using a small value of $\alpha$. Firstly, the dark states now have support on the vacuum states in both cavities, and so there is a chance we will declare failure when we prepare a dark state. This reduces our scheme's success probability to 
\begin{equation}
\label{eq:true_success_prob}
    p_{\text{success}}=\frac{1}{2}\left(1-2|\braket{0|\alpha}|^2+|\braket{0|\alpha}|^4\right)=\frac{1}{2}\left(1-2e^{-|\alpha|^2}+e^{-2|\alpha|^2}\right),
\end{equation}
where we have used the formula $P(A\cup B) = P(A) + P(B) - P(A\cap B)$. For large values of $\alpha$ this saturates to 1/2 as expected. 
This is the success probability when we perform the vacuum check by measuring both cavities and matches well with what we observe in Fig.\,\ref{fig:Fig3} in the main text. 
\\
\\
The other consequence is that our remaining dark state is altered by the projective measurement such that the state no longer has any chance of containing zero photons. If the initial dark state we prepare is denoted as $\ket{\text{dark}}$, then the state we ultimately end up with after passing the vacuum check would be
\begin{equation}
    \ket{\text{dark}'}\propto\hat{\Pi}_{\bar{0}\bar{0}}\ket{\text{dark}},
\end{equation}
where $\hat{\Pi}_{\bar{0}\bar{0}}$ is the `not-zero' projector
\begin{equation}
\hat{\Pi}_{\bar{0}\bar{0}}=(\mathds{1}-\ket{0}\bra{0})_1\otimes \mathds{1}_b\otimes (\mathds{1}-\ket{0}\bra{0})_2. 
\end{equation}
At first glance, removing vacuum would seem to reduce our entanglement fidelity, since it reduces our overlap with the target `cat-in-two-boxes' Bell state. However, by redefining the basis of our bosonic encoding, we can show $\ket{\text{dark}'}$ is a perfect Bell state in this basis. The modified basis is
\begin{align}
\label{eq:mod_basis}
    \begin{split}
        \ket{+}_L &\propto (\mathds{1}-\ket{0}\bra{0})\left(\ket{\alpha} + \ket{-\alpha}\right)\\
        \ket{-}_L &\propto \ket{\alpha} - \ket{-\alpha}
    \end{split}
\end{align}
where the $\ket{-}_L$ state is unchanged from the cat code but we subtract vacuum from $\ket{+}_L$. It is easy to check these states are orthogonal since $\ket{+}_L$ contains only even photon numbers whilst $\ket{-}_L$ contains only odd. In Fig.\,\ref{fig:Spheres2}. we show the Bloch sphere and Wigner functions for the modified basis and compare it to the standard cat code.
\\
\\
Now we show how we end up in a perfect Bell state if our target state was initially the dark state
\begin{equation}
\label{eq:best_ds}
    \ket{\text{dark}}\propto\ket{\alpha}_1\ket{-\alpha}_2-\ket{-\alpha}_1\ket{\alpha}_2
\end{equation}
which we can rewrite as 
\begin{align}
    \ket{\text{dark}}&\propto\left(\ket{\alpha}+\ket{-\alpha}\right)_1\left(\ket{\alpha}-\ket{-\alpha}\right)_2
    -\,\left(\ket{\alpha}-\ket{-\alpha}\right)_1\left(\ket{\alpha}+\ket{-\alpha}\right)_2
\end{align}
If we use the standard definition of the cat code 
\begin{equation}
        \ket{\pm}_\text{cat} = \frac{\ket{\alpha} \pm \ket{-\alpha}}{\sqrt{N_{\pm}}},
\end{equation}
where $N_{\pm}=2(1\pm e^{-2|\alpha|^2})$, then our initial dark state can be rewritten exactly as 
\begin{equation}    \ket{\text{dark}}=\frac{\ket{+}_\text{cat,1}\ket{-}_\text{cat,2}-\ket{-}_\text{cat,1}\ket{+}_\text{cat,2}}{\sqrt{2}}.
\end{equation}
It then follows that after applying the projector, $\hat{\Pi}_{\bar{0}\bar{0}}$ to this state we end up in the final dark state
\begin{equation}
\ket{\text{dark}'}=\frac{\ket{+}_\text{L,1}\ket{-}_\text{L,2}-\ket{-}_\text{L,1}\ket{+}_\text{L,2}}{\sqrt{2}}    
\end{equation}
which is indeed a perfect Bell state in the new basis. This is not true for all target dark states. For instance, if our target dark state was instead
\begin{equation}
    \ket{\text{dark}}\propto\ket{\alpha}_1\ket{-\alpha}_2+\ket{-\alpha}_1\ket{\alpha}_2,    
\end{equation}
then 
\begin{equation}    \ket{\text{dark}}\approx\frac{\ket{+}_\text{cat,1}\ket{+}_\text{cat,2}-\ket{-}_\text{cat,1}\ket{-}_\text{cat,2}}{\sqrt{2}},
\end{equation}
is not exact, owing to $N_+\neq N_-$, and the final entangled state after the vacuum check is not an exact Bell state in any basis. This is the main reason why we chose to initialize the cavities in the initial state
\begin{equation}
    (\ket{\alpha}+i\ket{-\alpha})_1\otimes(\ket{\alpha}-i\ket{-\alpha})_2
\end{equation}
which gives us the target dark state in Eq. \ref{eq:best_ds}.
\\
\\
Overall, this analysis shows that in order to produce a perfect Bell state at finite $\alpha$, we must modify the basis according to Eq. \ref{eq:mod_basis} and our target dark state must have odd joint photon number parity.
\\
\\
Interestingly, the modified basis DMM scheme still produces perfect entanglement even as $\alpha\rightarrow0$, provided that there is non-zero population in the $\ket{2}$ state. (The DMM scheme does not work for two-level systems). Although the success probability becomes vanishingly small, this limit shows the minimum Hilbert space size needed for the DMM scheme to theoretically succeed is three energy levels in each qubit. Also in this limit, $\ket{+}_L\rightarrow\ket{2}$ and $\ket{-}_L\rightarrow{\ket{1}}$ which begins to be apparent in the Wigner functions plotted in Fig.\,\ref{fig:Spheres2}.
Experimentally, we are willing to trade the success probability of our scheme by working with smaller $\alpha$ entangled states in order to obtain higher entanglement fidelities, since decoherence from photon loss is reduced when working at smaller mean cavity photon numbers.
\section{Single photon transfer efficiency}
\label{app:transfer_efficiency}
A useful figure of merit to help compare our scheme to other quantum communication schemes is the single photon transfer efficiency. We define this to be the probability of receiving a photon in module 2, after we first prepare a photon in module 1 and send it through the link. This quantity is often directly related to the quantum state transfer fidelity in deterministic schemes such as those described in ~\cite{Majer2007,Luke2020,2019ClelandRemote,2022ClelandEntanglementPurification,Zhong2023,Mollenhauer2025}. For our link, we quote a single photon transfer efficiency of 2.2\%, which implies these deterministic schemes would be completely unviable in our device. We arrive at this number by considering a simple state transfer scheme in which we actuate the beamsplitter coupling,
\begin{equation}
    \mathcal{H}_1/\hbar = \gbs ( a_1 b^\dagger + a_1^\dagger b ),
\end{equation}
for time $t_1$, which swaps the photon into the lossy bus mode. We then actuate the other beamsplitter coupling to swap the photon from the bus to the cavity in the other module, described by
\begin{equation}
        \mathcal{H}_2/\hbar = \gbs ( a_2 b^\dagger + a_2^\dagger b ),
\end{equation}
for time $t_2$. We set $\gbs/2\pi$ = \SI{160}{kHz} and $\kappa_b/2\pi$ = \SI{600}{kHz} and find the optimum values of $t_1$ and $t_2$ using QuTiP master equation simulations, from which we obtain $t_1=t_2=$\SI{1016}{ns} and ultimately an energy transfer efficiency of 2.2\%

\section{Is the DMM scheme really completely robust to bus loss?}
\subsection{Off-resonant couplings to other cable harmonics}
\label{app:secret_loss}

\noindent At first glance, it appears that the DMM scheme completely avoids all loss in the bus mode, for successful attempts.
This is true when the channel can be accurately modelled as a single standing wave mode. 
However, any finite-length link must necessarily support multiple standing wave modes, with a frequency spacing dictated by $\fsr$.
It is thus impossible to only couple to one of these standing wave modes, without off-resonantly coupling to the other standing wave modes. 
 In our circuit-QED hardware, $
\fsr$ exceeds $\gbs$ by approximately three orders of magnitude, so this effect is extremely small. Nevertheless, it is insightful to see how these these off-resonant couplings that present a fundamental source of loss in the DMM scheme.
\\
\\
To see how, suppose we engineer a beamsplitter coupling between our cavity qubits and the $n^{\textrm{th}}$ harmonic of the cable link. We will also couple off-resonantly to the $(n+1)^{\textrm{th}}$ and $(n-1)^{\textrm{th}}$ harmonic. We can write a more accurate model for our coupling Hamiltonian as
\begin{align}
\begin{split}
    \mathcal{H}_{c}/\hbar &= g_{\text{BS}}(a_1 + a_2)b_n^\dagger\\
    &+ g_{\text{BS}}(a_1 - a_2)b_{n+1}^\dagger + \Delta_{\text{FSR}} b_{n+1}^\dagger b_{n+1}\\
    &+ g_{\text{BS}}(a_1 - a_2)b_{n-1}^\dagger - \Delta_{\text{FSR}} b_{n-1}^\dagger b_{n-1}\\
    &+ h.c.
\end{split}
\end{align}
where $b_n$ is the $n^{\textrm{th}}$ harmonic of the cable and $\Delta_{\text{FSR}}$ is the Free Spectral Range. 
Although we have only accounted for the first two off-resonant couplings, we can already see how this presents an additional source of loss. 
Suppose we start in a dark mode such as $\ket{\alpha}_1\ket{-\alpha}_2$. Whilst this is truly a dark mode for the coupling to mode $b_n$, it is in fact a bright mode for the off resonant couplings to $b_{n+1}$ and $b_{n-1}$.
This is owing to the fact that neighbouring harmonics are of opposite parity, alternating between even and odd spatial field profiles , and manifests as a sign change between $a_1$ and $a_2$ for the off-resonant couplings.
\\
\\
How does this loss scale with the physical parameters? If we assume each harmonic has the same energy loss rate, $\kappa_b$ and a time to generate entanglement $\sim 1/g_{\text{BS}}$ for successful attempts, then we find the total fraction of energy lost in the other cable harmonics to be 
\begin{align}
\begin{split}    \varepsilon_{\text{loss}}&=2\left(\frac{\kappa_b}{g_{\text{BS}}}\right)\left[\left(\frac{g_{\text{BS}}}{\Delta_{\text{FSR}}}\right)^2+\left(\frac{g_{\text{BS}}}{3\Delta_{\text{FSR}}}\right)^2+...\right]\\
&\approx 2\,\left(\frac{g_{\text{BS}}}{\Delta_{\text{FSR}}}\right)\left(\frac{\kappa_b}{\Delta_{\text{FSR}}}\right).
\end{split}
\end{align}

This loss primarily reduces fringe visibility (the error associated with the incorrect joint photon number) for the cat-in-2-boxes state and introduces an infidelity given by
\begin{equation}
    \bar{F}_\text{loss} = 2|\alpha|^2\varepsilon_\text{loss},
\end{equation}
\\
which is essentially just the probability a single photon loss occurs during the entanglement scheme. 
For our system, this estimated to be  $\sim10^{-6}$ and is far from being the limiting contribution to the entanglement fidelity. 
This contribution becomes more significant if $\Delta_{\text{FSR}}$ approaches $g_{\text{BS}}$.
\\
\\
Another way to think about this loss mechanism is that there is a finite timescale, set by the speed of light in the cable, that is required for destructive interference to occur. In other words, transient wavepackets are excited from both ends of the cable until they destructively interfere with each other to form effervescent waves. Since all of this happens on a timescale set by $\Delta_{\text{FSR}}$, and the wave-packet amplitudes are small, this results in a small source of energy loss.  One thing to note is that this loss can be much lower than the single pass loss (SPL), the minimum energy loss a photon experiences when `travelling' from one end of the link to the other. 
This is given by 
\begin{equation}
    \varepsilon_{\text{SPL}}=\kappa_b \frac{L}{c}=\frac{\kappa_b}{2\Delta_{\text{FSR}}}.
\end{equation}
Note that $\varepsilon_{\text{loss}}<\varepsilon_{\text{SPL}}$ is only possible because our scheme is probabilistic. Similarly, in optical heralded entanglement schemes, the entanglement fidelity routinely exceeds the single pass loss of the channel. 
However, any deterministic communication scheme, including adiabatic schemes, is fundamentally limited by the single pass loss, an observation also noted by\,\cite{2019SchusterDarkMode}. 
In our system, we estimate the single pass loss to be $\sim10^{-4}$. 

For this particular cable, the FSR is around 2 GHz. Harmonics have been measured at \SI{3.834}{GHz}, \SI{5.658}{GHZ}, \SI{7.336}{GHz} and \SI{8.927}{GHz}. The harmonics are not exactly evenly spaced in frequency, due to the varying participations of each harmonic in the Teflon dielectric. 
\subsection{Purcell limit from a lossy bus mode}
\noindent As noted in Sec. \ref{sec:scheme}, our scheme does not protect from photon loss errors in the cavities, which is thought to be the dominant error source. The cavity mode weakly hybridizes with both the bare transmon mode and the bare cable modes. If the modes in the cable are too lossy, we will observe a reduction in cavity $T_1$ due to the (inverse) Purcell effect. Furthermore, depending on the value of $\Delta_{\text{FSR}}$ the multimode Purcell effect can also be important. Here we again estimate the scaling of this error source with system parameters and show that for our system there is a larger contribution than the off-resonant couplings but still too small to be measured compared to the other sources of errors. 
\\
\\
For this, we estimate what fraction of a dressed cavity mode participates in the bare lossy bus mode. We assume that direct coupling between the bare cavity mode and cable mode is negligible compared to the hybridization enabled by both the cavity and cable modes participating in the bare transmon mode. The energy participation in the bare transmon mode can be estimated from the measured $\chi$'s and the anharmonicities of the transmons, $\alpha_t$. We thus expect each cavity to inherit a loss rate from the bus mode given by
\begin{equation}
    \kappa_\text{cav}^{\text{Purcell}}=\frac{\chi_{a_it}}{\alpha_t}\frac{\chi_{bt}}{\alpha_t}\kappa_b
\end{equation}
which contributes 
\begin{equation}
    \bar{F}_{\text{Purcell}}=2|\alpha|^2\frac{\kappa_\text{cav}^{\text{Purcell}}}{g_{\text{BS}}},
\end{equation}
to the entanglement infidelity. For our system parameters we estimate this contribution to be $\sim10^{-4}$. In general, this effect can be reduced if needed by reducing the participation of cavity and cable modes in the bare transmon mode whilst maintaining the same $g_{\text{BS}}$ rate (i.e. decrease $\chi_{a_it}$ and/or $\chi_{bt}$ and compensate with stronger pumps). If we are in an experimental regime where a smaller $\Delta_{\text{FSR}}$ results in multiple cable harmonics having an appreciable $\chi$ with the transmon, then the multimode Purcell effect may further limit cavity $T_1$. In this experiment, we do not expect the cavities to be Purcell-limited by the transmons with their lifetimes set to $T_1^{\text{Purcell}}>$\SI{1}{ms}. Instead, we attribute the dominant error in our Bell state generation to photon loss in the cavities during the DMM scheme and errors introduced by our tomography step. (See ~\ref{app:budget})

\section{High-fidelity cat state preparation}
\label{app:cat_prep}
\begin{figure*}[h] 
\includegraphics[scale = 0.5]{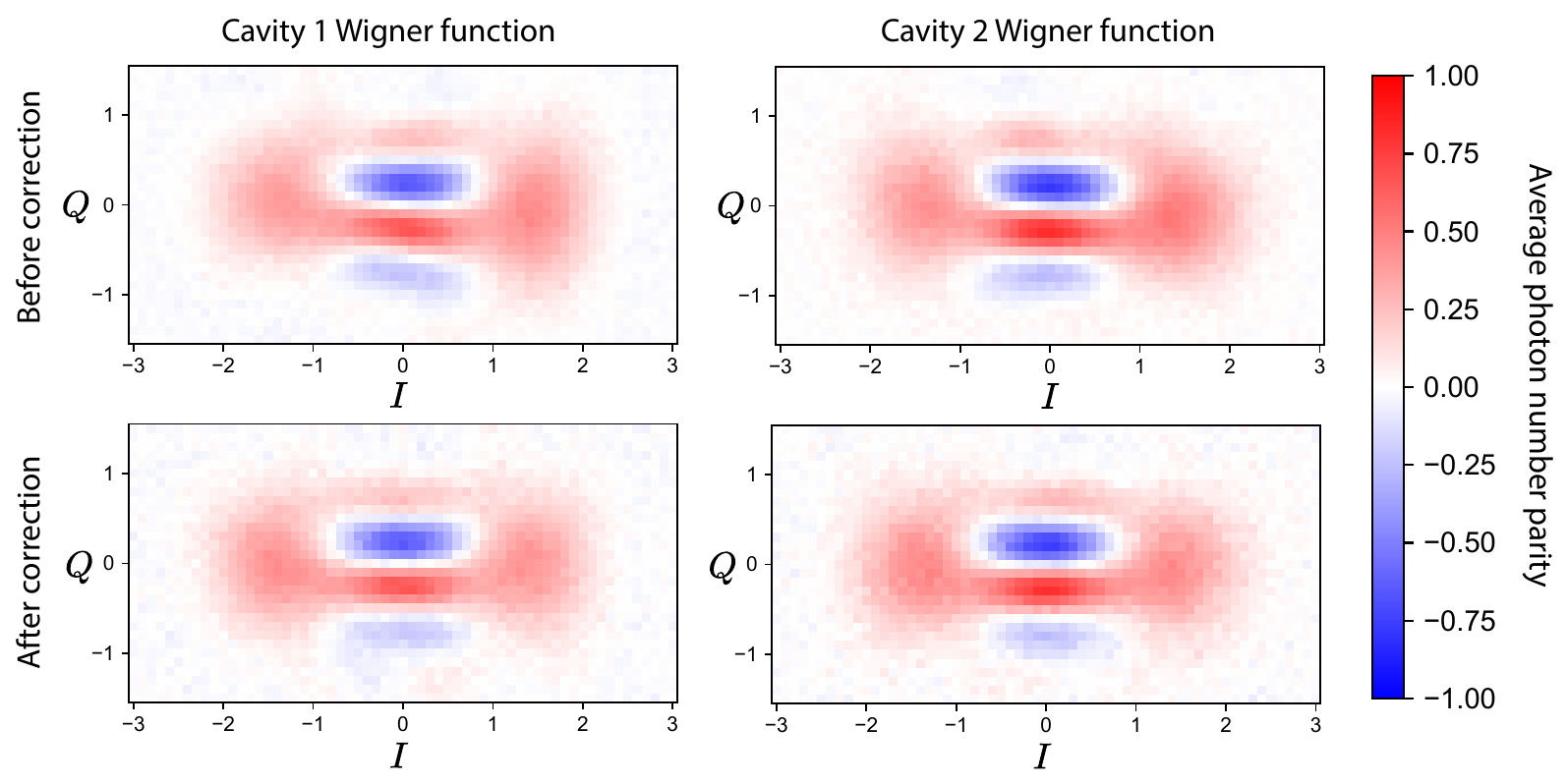}
\caption{\textbf{Correcting for pulse distortions in the control lines during cat state preparation.} We use an OCT SNAP pulse to prepare the cat state $\propto\ket{\alpha}+i\ket{\alpha}$ after initializing in $\ket{\alpha}$, where $\alpha=1.414$. If we initialize in $\ket{-\alpha}$ instead, we prepare the state $\propto\ket{\alpha}-i\ket{\alpha}$. From the measured Wigner functions we see these states are initially slightly distorted (top row), which we attribute to unknown transfer functions such as dispersion in the control lines. After reconstructing the density matrix to find $\tilde{\mathcal{U}}$, we re-generate the OCT pulse and measure the Wigner function again (bottom row), where we see less distortion in the fringes, especially for cavity 1. We could iterate over this process many times but we find going through this process once suffices to prepare cat states in excess of 99\% state fidelity.}
\label{fig:cat_dist}
\end{figure*}

For the DMM scheme to give the best Bell state fidelity at finite $\alpha$, our target state must have odd joint photon number parity. Our starting states can either be an odd or even cat, for example
\begin{equation}
    \ket{\psi_{\text{init}}} \propto (\ket{\alpha}+\ket{-\alpha})_1\otimes\ket{0}_b\otimes(\ket{\alpha}-\ket{-\alpha}),
\end{equation}
which can be prepared relatively quickly in $\sim$~\SI{1}{\micro\second} via optimal control pulses on the cavity and transmon simultaneously. Alternatively, we could begin from so-called `parity-less' cat states of the form
\begin{equation}
    \ket{\psi_{\text{init}}} \propto (\ket{\alpha}+i\ket{-\alpha})_1\otimes\ket{0}_b\otimes(\ket{\alpha}-i\ket{-\alpha}).
\end{equation}
We opt for this latter approach, because we can prepare `parity-less' cats with a higher fidelity, estimated to be $\sim99\%$ using the procedure we outline below.
\\
\\
First we use simple displacement pulses to prepare $\ket{\alpha}$ in both cavities. We then apply an optimal control pulse to each transmon that functions similarly to a SNAP pulse. For example, the unitary that takes $\ket{\alpha}\rightarrow\ket{\alpha}\pm i\ket{-\alpha}$ can be written as a SNAP unitary of the form: 
\begin{equation}
    \mathcal{U}_{\text{SNAP}} = \sum_{n=0}^\infty e^{i\theta_n}\ket{n}\bra{n},
\end{equation}
where $\theta_n = \pi/4$ for $n = \text{even}$ and $\theta_n = -\pi/4$ for $n = \text{odd}$. Whilst we can construct this SNAP pulse by hand, we find OCT pulses that are constrained to only drive the transmon perform better, since effects such as the finite selectivity of $\pi$-pulses and cavity self-Kerr are automatically accounted for. As for a regular SNAP pulse, the OCT pulse leaves the transmon back in its ground state at the end of the pulse. 
\\
\\
When we implement these `OCT-SNAP' pulses in experiment, we see visible distortions in the Wigner functions, typically in the fringes or `whiskers' of the cat. We attribute this to frequency dispersion in our control lines which distorts our OCT pulse. This means the actual state we prepare is of the form:
\begin{equation}
    \ket{\tilde{\psi}}=\tilde{\mathcal{U}}(\ket{\alpha}\pm\ket{-\alpha})/\sqrt{2},
\end{equation}
where $\tilde{\mathcal{U}}$ is an unwanted SNAP unitary, also of the form $\tilde{\mathcal{U}}=\sum_{n=0}^\infty e^{i\tilde{\theta}_n}\ket{n}\bra{n}$ where $\tilde{\theta}_n$ are unwanted phase distortions due to the unknown transfer function of the control lines. 
However, since we can reconstruct the density matrix from the Wigner function, we can find the values of $\tilde{\theta}_n$ from the density matrix. To correct for this effect, we simply ask our OCT optimization to create a pulse that implements the unitary $\tilde{\mathcal{U}}^\dagger\mathcal{U}_{\text{SNAP}}$ instead, using the previous pulse as the seed waveform. We find this improves the appearance of the fringes and increases state fidelity by 1--2\%. Fig.\,\ref{fig:cat_dist}. shows the Wigner functions of the prepared cat states before and after this correction.
\newpage
\section{Extended data sets for Bell state generaton.}
For each point in the Wigner function, $W(\beta)$ we take 10,000 experimental shots, before post-selecting on the vacuum check outcomes. The exception to this is at $\alpha = 1$ where we take 4,000 experimental shots, resulting in a larger error bar in the Bell state fidelity. The resolution of the Wigner function grid is 0.1 in phase space displacements, $\beta$.

\begin{figure*}[h] 
\label{app:ent_ext}
\includegraphics[scale = 0.5]{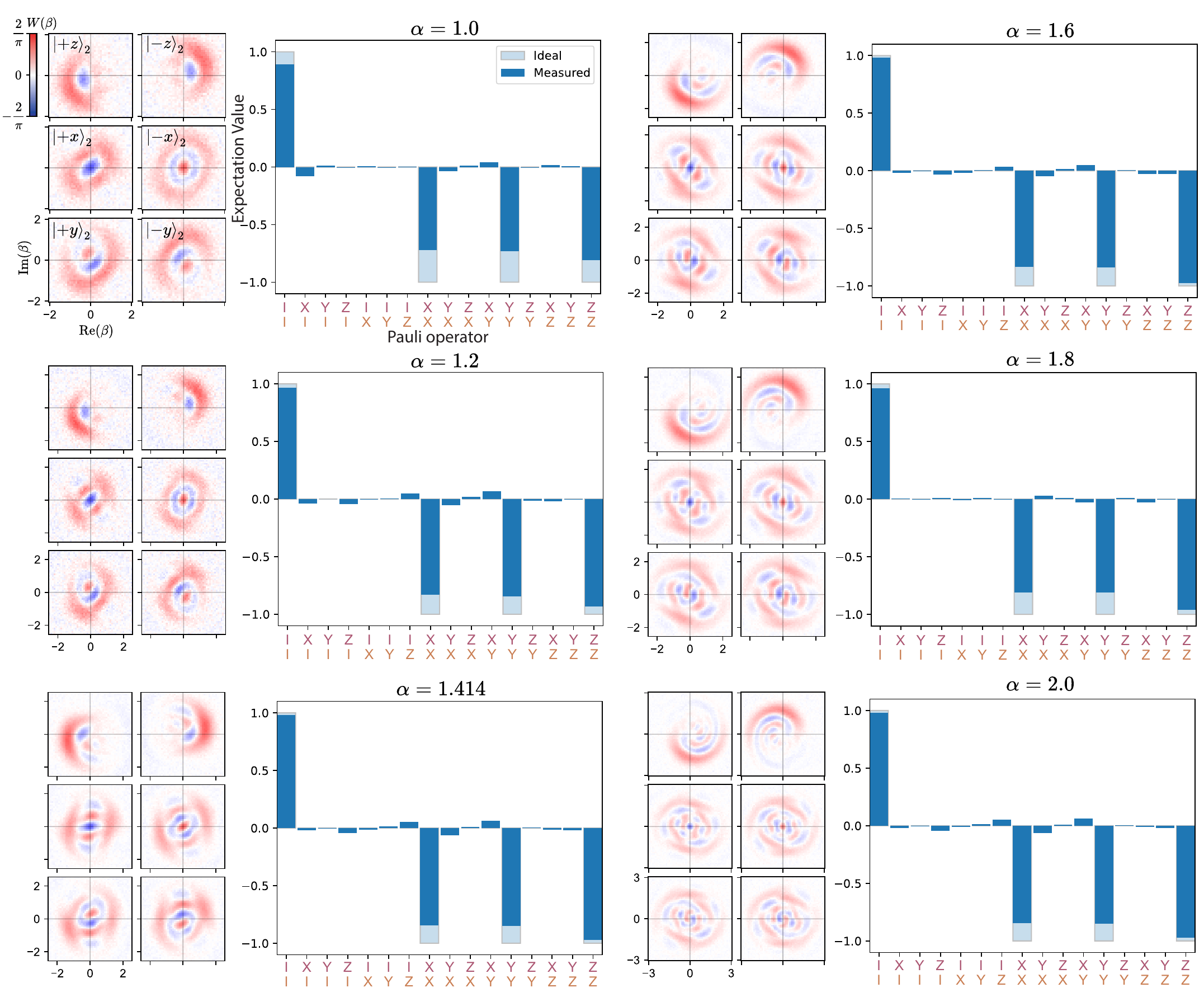}
\caption{\textbf{Extended data set for Bell state generation}. For various values of $\alpha$, we show the measured symmetrized Wigner functions alongside the logical two-qubit Pauli correlators from which the Bell state fidelity is calculated. No active corrections to Kerr distortion are applied to this data. Instead, we update the logical basis of cavity 1 accordingly. Kerr distortions become visibly more noticeable as $\alpha$ increases. The $ZZ$ correlation is consistently greater than the $XX$ or $YY$ correlations, suggesting our dominant error is cavity photon loss and that we have excellent distinguishability between bright and dark states. The $II$ correlation is slightly less than 1 and represents leakage out of cavity 1's logical codespace, predominantly into the vacuum state due to photon loss. Small correlations in $XY$ or $YX$ are thought to be due to small phase miscalibrations in the decode pulses, but have minimal impact on the Bell state fidelity ($<0.5\%$)}
\label{fig:ent_bell_full}
\end{figure*}

We numerically optimize the definition of our logical basis to maximize the Bell state fidelity. This basis is parameterized by three parameters: $\alpha_{\text{basis}},\theta_K,\theta_r$ which respectively account for shrinking $\alpha$ due to the amplitude damping back-action associated with cavity photon loss, Kerr distortion during the time of the protocol and any miscalibration of drive phases that leads to a cavity rotation offset. 
The logical basis is defined from these parameters as:
\begin{align}
\begin{split}
\label{eq:param_basis}
    \ket{+}_L \propto e^{i\theta_r \hat{a}^\dagger \hat{a}}e^{\frac{i}{2} \theta_K \hat{a}^\dagger \hat{a}^\dagger \hat{a} \hat{a}}\hat{\Pi}_{\bar{0}}( \ket{\alpha_{\text{basis}}} &+ \ket{-\alpha_{\text{basis}}})\\
    \ket{-}_L \propto e^{i\theta_r \hat{a}^\dagger \hat{a}}e^{\frac{i}{2} \theta_K \hat{a}^\dagger \hat{a}^\dagger \hat{a} \hat{a}}( \ket{\alpha_{\text{basis}}} &- \ket{-\alpha_{\text{basis}}}),
\end{split}
\end{align}
where $\hat{\Pi}_{\bar{0}} = \mathds{1}-\ket{0}\bra{0}$.
We use this basis to construct the expectation values of joint-Pauli correlations (Pauli bars).
From the Pauli bars we can extract the $4\times4$ density matrix, $\rho_{L}$ that represents the entangled state in the logical basis. The Bell state fidelity is calculated from
\begin{equation}
    \mathcal{F}_{\text{Bell}}=\braket{\Psi_-|\rho_L|\Psi_-},
\end{equation}
where 
\begin{equation}
    \ket{\Psi_-} = \frac{\ket{+}_L\ket{-}_L - \ket{-}_L\ket{+}_L}{\sqrt{2}}.  
\end{equation}
The measured Wigner functions and extracted Pauli bars are shown in Fig.\,\ref{fig:ent_bell_full}, with quantitative metrics such as Bell state fidelity and success probabilities presented in Table \,\ref{tab:basis_parameters}.

\begin{table}[H]
\centering
\begin{tabular}{c|c|c|c|c|c|c|c}

\toprule  
\specialcell[t]{}    & $\alpha = 1.0$ & $\alpha = 1.2$ & $\alpha = 1.414$ & \specialcell[t]{$\alpha= 1.414$ \\ Kerr-cancelled} & $\alpha= 1.6$ & $\alpha = 1.8$ & $\alpha= 2.0$ \\
\hline
$\alpha_{\text{basis}}$    & 0.938& 1.125& 1.331 & 1.344& 1.508& 1.706& 1.904         \\
$\theta_K$  & 0.467& 0.473& 0.493 & 0.001& 0.502& 0.502& 0.502         \\
$\theta_r$  & 0.001& 0.010& -0.749& 0.009& 0.119& -0.012& -0.014        \\
\hline
\specialcell[t]{Success probability (\%)}  & 17.7& 26.3& 34.7 & \textbf{34.9}& 40.0& 44.2& 47.2         \\
\hline
\specialcell[t]{Bell state fidelity (\%)} & 79.0& 89.5& 91.3 & \textbf{92.3}& 90.8& 88.6& 85.6 \\
\toprule
\end{tabular}
\caption{\textbf{Results for Bell state generation via the DMM scheme, for different values of $\alpha$ used in the experiment.} We also report the optimum basis parameters, along with the overall success probability and Bell state fidelity.}
\label{tab:basis_parameters}
\end{table}
\newpage
\section{Two-qubit tomography}
\label{app:twobit_tomo}
In this section we show how we perform two-qubit tomography on the two cavities at the end of the DMM scheme, which allows us to extract joint Pauli correlations and thus reconstruct the state in its logical basis.
\subsection{Measuring a cavity in the logical basis}
We can directly measure a cavity mode in its logical basis by first mapping from its bosonically encoded logical states onto its transmon, and then measuring the transmon. This mapping is accomplished through an OCT pulse we call the decode pulse which takes
\begin{align}
    \begin{split}        \ket{0_L}\ket{g}&\rightarrow\ket{\text{vac}}\ket{g}\\    \ket{1_L}\ket{g}&\rightarrow\ket{\text{vac}}\ket{e}
    \end{split}
\end{align}
Reading out the transmon immediately after the decode pulse amounts to measuring the cavity in its Z logical basis. To measure in the $X_L$ or $Y_L$ basis, we use the same decode pulse but instead perform a $\pi/2$-pulse about the Y or X axis of the transmon before it is read out. We opt to measure cavity 2 this way (mapping its state onto its transmon), since its transmon has better coherence than the transmon of cavity 1.
\\
\\
It is important that we specify the correct logical basis $\{\ket{0_L},\ket{1_L}\}$ for the OCT pulse. Whenever we wish to decode cavity 2, we measure its Wigner function to determine the value of $\alpha$ we should use for the basis (which is always less than our starting $\alpha$ due to amplitude damping in the cavities), along with other unitary corrections such as rotations ($e^{i\theta_r \hat{a}^\dagger \hat{a}}$) and Kerr (($e^{i\frac{\theta_K}{2} \hat{a}^\dagger \hat{a}^\dagger \hat{a} \hat{a}}$)). This also means there is no need to actively correct Kerr distortions in cavity 2 before decoding, so long as we know the value of $\theta_K$ to use in the decode pulse.
\\
\\
The decode pulses we use take \SI{800}{ns} and are generated again for every value of $\alpha$ we test. We can estimate the error associated with these pulses by also generating another pulse that performs the inverse state transfer:
\begin{align}
    \begin{split}        \ket{\text{vac}}\ket{g}&\rightarrow\ket{0_L}\ket{g}\\    \ket{\text{vac}}\ket{e}&\rightarrow\ket{1_L}\ket{g},
    \end{split}
\end{align}
which we call the encode pulse. We prepare the six cardinal states in the transmon, perform the encode pulse and then immediately the decode pulse and then finally the transmon rotation that should put the transmon back in its ground state. Averaging over the cardinal states we find a 4\% probability we do not end up in the ground state and hence estimate a 2\% attributed to the decode pulse. We are willing to take this extra penalty in Bell state fidelity (due to this part of our tomography) since it circumvents the need to measure the joint Wigner function in order to characterize our state.
\subsection{Two-qubit state reconstruction}
The goal is to ultimately reconstruct the two-qubit density matrix of the entangled cavity states in their logical basis. The raw data we must process are the six filtered Wigner functions of cavity 1. We call this $\rho_{\pm l}^{(1)}$ for $l\in\{x,y,z\}$. Each Wigner function corresponds to a choice of logical basis and measurement outcome on cavity 2. In addition, we also know the probabilities of obtaining these outcomes which we denote as $p_{\pm}^{(2)}$.
\\
The first step is to reconstruct the density matrix of the state in cavity 1, in the Fock basis. We use Maximum Likelihood Estimation (MLE) reconstruction, truncating at 10 oscillator levels (13 for $\alpha = 2.0$). We also normalize the Wigner function data to 1 before performing MLE, which ensures $\text{Tr}\{\rho_{\text{recon}}\}=1$ thus corresponding to a physical state in the cavity. This corrects for uniform (non $\beta$-dependent) contrast reduction in the Wigner data (e.g. due to transmon decoherence in the parity measurement). We also benchmark our MLE reconstruction by generating simulated Wigner function data for state $\rho_{\text{test}}$, adding the expected shot noise (10,000 samples per point), using MLE to find $\rho_{\text{recon}}$, and comparing with $\rho_{\text{test}}$. From bootstrapping resampling we attribute an uncertainty of $0.3\%$ in the Bell state fidelity due to shot noise, but quote a 1\% error associated with our MLE reconstruction, attributed to the various systematic errors such as $\beta$-dependent contrast of $W(\beta)$ measurements and our sensitivity to any miscalibration of the $\text{Re}(\beta)$ and $\text{Im}(\beta)$ axes.
\\
\\
Once we choose a parameterized logical basis (see Eq. \ref{eq:param_basis}), we can calculate the expectation values of the logical joint Pauli operators, $\braket{\sigma_i\sigma_j}$ for $i,j \in \{I,X,Y,Z\}^{\otimes 2}$. Where $i$ $(j)$ is the index for cavity 1 (2). We can also define the set of logical Pauli operators written in the Fock basis of cavity 1 as $\sigma_i^{(1)}\in\{I_L^{(1)},X_L^{(1)},Y_L^{(1)},Z_L^{(1)}\}$For the case where $j\neq I_L$, the joint Pauli expectation values are calculated as 
\begin{equation}
\braket{\sigma_i\sigma_j} = p_{+j}^{(2)}\text{Tr}\left(\rho_{+j}^{(1)}\sigma_i^{(1)}\right)-p_{-j}^{(2)}\text{Tr}\left(\rho_{-j}^{(1)}\sigma_i^{(1)}\right)
\end{equation}
When $j=I_L$ we must modify this to be
\begin{equation}
\braket{\sigma_i\sigma_I}= p_{+j}^{(2)}\text{Tr}\left(\rho_{+j}^{(1)}\sigma_i^{(1)}\right)+p_{-j}^{(2)}\text{Tr}\left(\rho_{-j}^{(1)}\sigma_i^{(1)}\right),   
\end{equation}
and similarly for $\braket{\sigma_I\sigma_j}$
It does not matter which basis, $j$ we use to calculate these single-qubit Pauli correlations, $\braket{\sigma_i\sigma_j}$ but in our case, we choose to average over all three.
\\
\indent Once we have calculated all 16 expectation values, we can directly reconstruct the two-qubit density matrix. However, due to measurement errors, there is no guarantee that the density matrix is physical. We then use MLE again, this time to find the closest physically allowed density matrix that best fits this directly reconstructed density matrix. This time, we do not constrain this MLE density matrix to be normalized (since we can now have leakage outside the logical basis, which is a subspace of the full cavity Hilbert space) but it must still be Hermitian and positive semidefinite. It is from this final density matrix that we quote our Bell state fidelity. 
\\
In order to find the optimum logical basis and Bell state fidelity, we numerically optimize a function that calculates the Bell state fidelity for a given a set of parameters that parametrize the logical basis. In each function evaluation we calculate the new Pauli bars and perform the MLE reconstruction.
\section{Calculating State Teleportation Fidelity}
\label{app:tele_processing}
Teleportation fidelity is calculated as a quantum state transfer (QST) fidelity. In this sense, we can calculate the average QST fidelity of an unknown state from the transfer fidelity of the 6 cardinal qubit states $\ket{0},\ket{1},\ket{+},\ket{-},\ket{+i},\ket{-i}$. The transfer fidelity for the states $\ket{+}$ and $\ket{-}$ will be equal, since these states only differ by a phase space rotation in the cat code, and similarly for the $\ket{\pm i}$ states. This means the average state transfer fidelity of an unknown qubit state is given by 
\begin{equation}
    \mathcal{F}_{\text{QST}}^\text{avg}=\frac{\mathcal{F}_{\text{QST}}^{\ket{0}}+\mathcal{F}_{\text{QST}}^{\ket{1}}+2\mathcal{F}_{\text{QST}}^{\ket{+}}+2\mathcal{F}_{\text{QST}}^{\ket{+i}}}{6},
\end{equation}
which we use to calculate the average state transfer fidelity for our teleportation channel. After we have chosen a parameterized logical basis for cavity 1, we calculate the state transfer fidelity of known qubit state $\ket{\psi}$ as
\begin{equation}
    \mathcal{F}_{\text{QST}}^{\ket{\psi}} = \text{P}(0,0)\braket{\psi_L|\rho_{00}^{\text{meas}}|\psi_L}+\text{P}(0,1)\braket{\psi_L|X_L\rho_{01}^{\text{meas}}X_L|\psi_L}+\text{P}(1,0)\braket{\psi_L|Z_L\rho_{10}^{\text{meas}}Z_L|\psi_L}+\text{P}(1,1)\braket{\psi_L|Y_L\rho_{01}^{\text{meas}}Y_L|\psi_L},
\end{equation}
where $X_L,Y_L,Z_L$ are the logical Pauli operators written in the cavity 1 basis and amount to performing the Pauli corrections in software. $\text{P}(m_1,m_2)$ are the outcomes of the bell measurements (all close to 1/4) and $\rho_{m_1,m_2}$ are the MLE reconstructed density matrices of cavity 1 obtained from the measured Wigner functions. 
\\
\\
We numerically optimize $\mathcal{F}_{\text{QST}}^\text{avg}$ with respect to the choice of logical basis for cavity 1 which is once again parameterized by three parameters $\alpha_{\text{basis}},\theta_r,\theta_K$ according to Eq. \ref{eq:param_basis}. The parameters used are reported in Tab. \ref{tab:Kerred_parameters}
\newpage
\section{Extended data sets for quantum state teleportation}
\begin{figure*}[h] 
\includegraphics[scale = 0.75]{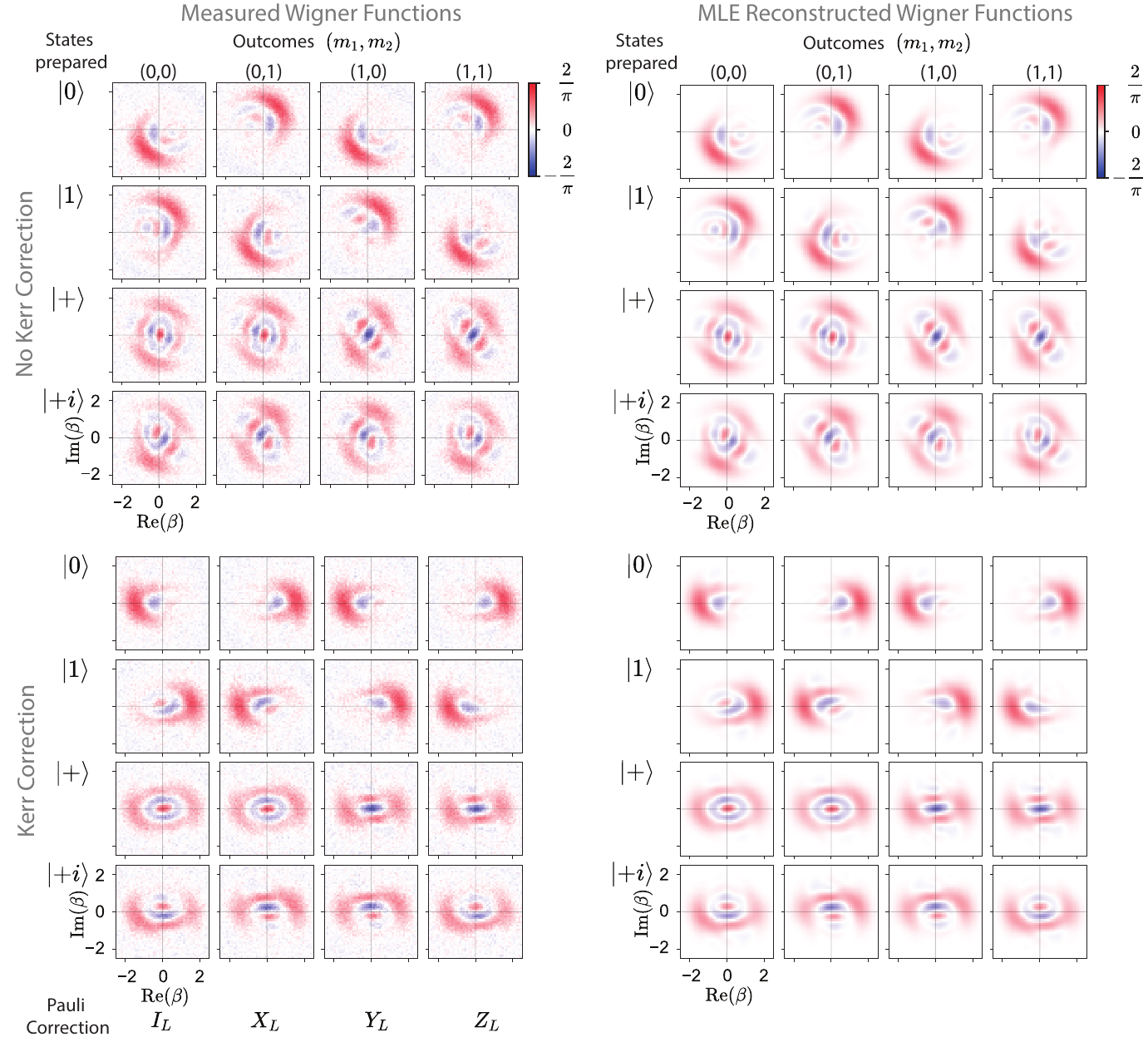}
\caption{\textbf{Measured and reconstructed Wigner functions for state teleportation, with and without Kerr compensation.} (Top left) Measured Wigner functions for the state teleported to cavity 1 without the Kerr-compensating pulse applied. This is the data from which we quote $90\pm1\%$ state transfer fidelity. (Top right) b) Wigner functions after performing maximum likelihood estimation (MLE) reconstruction. From a measured Wigner function, we use MLE to find the density matrix of the cavity state whose Wigner function (as shown) best matches the measured Wigner functions on the left. (Bottom Left) Measured Wigner functions for the state teleported to cavity 1 with the Kerr-compensating pulse applied. This is the data from which we quote $89\pm1\%$ state transfer fidelity. Visually it is more obvious which Pauli corrections would need to be applied to cavity 1. Qualitatively, we can see the Kerr-compensating pulse still leaves some visible Kerr distortions and a deterministic rotation offset. (Bottom right) Reconstructed Wigner functions for the data set to the left.}
\label{fig:ent_tele_full}
\end{figure*}
\begin{table}[h]
\centering
\begin{tabular}{ccl|c|c|c|c|cc}
\toprule
\\
 Kerr correction?&Metric&State prepared& \multicolumn{4}{c|}{Outcome $(m_1,m_2)$}& \\

 &&& (0,0) & (0,1) & (1,0) & (1,1) & Average \\
\hline
No&Fidelity&$\ket{0}$ & 0.9561 & 0.9644 & 0.9506 & 0.9142 & 0.9463 \\
&&$\ket{1}$ & 0.9000 & 0.9142 & 0.9206 & 0.9011 & 0.9090 \\
&&$\ket{+}$ & 0.8915 & 0.8881 & 0.990 & 0.9190 & 0.9094 \\
&&$\ket{+i}$ & 0.8991 & 0.8673 & 0.8693 & 0.8527 & 0.8721\\
&$\mathcal{F}_{\text{QST}}^\text{avg}$&&&&&&$90\pm1\%$&\\
&&&&&&&&\\
&Probability&$\ket{0}$ & 0.2485 & 0.2493 & 0.2516 & 0.2506 \\
&&$\ket{1}$ & 0.2508 & 0.2509 & 0.2496 & 0.2487 \\
&&$\ket{+}$ & 0.2491 & 0.2501 & 0.2521 & 0.2487 \\
&&$\ket{+i}$ & 0.2494 & 0.2497 & 0.2507 & 0.2501\\
&&&&&&&&\\
Yes&Fidelity&$\ket{0}$ & 0.9443 & 0.9681 & 0.9387 & 0.8981 & 0.9372 \\
&&$\ket{1}$ & 0.8724 & 0.8990 & 0.8985 & 0.9102 & 0.8950 \\
&&$\ket{+}$ & 0.8915 & 0.8881 & 0.9274 & 0.9079 & 0.8998 \\
&&$\ket{+i}$ & 0.8598 &  0.8608 & 0.8749 & 0.8472 & 0.8607\\
&$\mathcal{F}_{\text{QST}}^\text{avg}$&&&&&&$89\pm1\%$&\\
&&&&&&&&\\
&Probability&$\ket{0}$ & 0.2483 & 0.2495 & 0.2514 & 0.2507 \\
&&$\ket{1}$ & 0.2509 & 0.2506 & 0.2500 & 0.2485 \\
&&$\ket{+}$ & 0.2491 & 0.2501 & 0.2521 & 0.2486 \\
&&$\ket{+i}$ & 0.2494 & 0.2498 & 0.2508 & 0.2500\\
&&&&&&&&\\
\toprule
\end{tabular}
\caption{\textbf{Quantum state transfer via teleportation}. Teleportation fidelities and success probabilities for four of the cardinal states with and without Kerr correction. From this, we can infer the average state transfer of an unknown quantum state, denoted by $\mathcal{F}_{\text{QST}}^\text{avg}$.}
\label{tab:teleportation_fidelities}
\end{table}
\begin{table}[]
\begin{tabular}{c|c|c|c|ccc|c|c|c}

\toprule 
\textbf{}&\multicolumn{9}{c}{Bell measurement outcome $(m_1,m_2)$}\\
\textbf{}& \multicolumn{4}{c}{no Kerr-compensation}&&\multicolumn{4}{c}{Kerr-compensation}\\
   & $(0,0)$ & $(0,1)$ & $(1,0)$ & $(1,1)$ &\,& $(0,0)$ & $(0,1)$ & $(1,0)$ & $(1,1)$ \\
\hline
$\alpha_{\text{basis}}$    & 1.319& 1.327& 1.319 & 1.321&& 1.317& 1.325& 1.317 & 1.313        \\
$\theta_K$  & 0.515& 0.509& 0.474 & 0.471&& -0.058& -0.072& -0.138 & -0.129        \\
$\theta_r$  & -0.012& 0.000& 0.113& 0.123&& 0.198& 0.226& 0.381& 0.366       \\

\toprule
\end{tabular}
\caption{\textbf{Logical Basis parameters for the teleportation data set.} Basis parameters for the different outcomes of the Bell state measurements, optimised for with and without the Kerr compensation pulse.}
\label{tab:Kerred_parameters}
\end{table}

\section{3D Purcell filter for the bus mode}
\label{app:Purcell}
\begin{figure*}[h] 
\includegraphics[scale = 1.2]{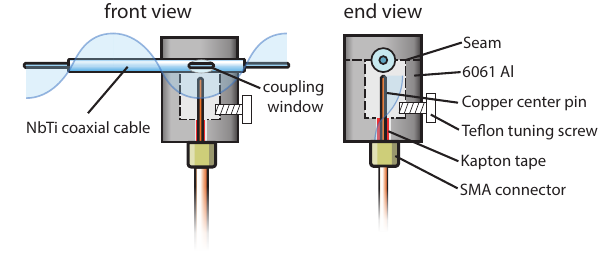}
\caption{\textbf{Design of a 3D Purcell filter for the bus mode.} A window is drilled through the outer conductor of the bus to expose the inner conductor. The Purcell filter mode is the $\lambda/4$ mode of a microwave stub cavity resonator, which capacitively couples to the bus mode. The frequency of this mode is set by the length of the copper center pin but can be fine-tuned via the Teflon tuning screw. The copper center pin directly protrudes from an SMA flange connector, meaning the purcell filter mode is galvanically coupled to the 50\,$\Omega$ transmission line. However, the copper pin passes through a narrow tunnel in between the stub cavity and the SMA flange. The sudden change in tunnel width creates an impedance mismatch that boosts the Quality factor of the Purcell filter mode. Kapton tape prevents the copper center pin from shorting with the 6061 Aluminum housing.}
\label{fig:purcell}
\end{figure*}

\noindent Why do we need to increase the loss of the bus mode beyond its intrinsic loss rate? The answer is related to the details of how we are performing our vacuum check. In general, the DMM scheme always benefits from a faster beamsplitter rate, $\gbs$, which sets the speed at which we can generate entanglement on a given attempt.  
Engineering larger participation of the cavity and bus mode in the transmon's Josephson junction helps us achieve fast beamsplitter rates at modest pump amplitudes but results in an appreciable dispersive interaction not only between the transmon and the cavity, but also between the transmon and the bus mode. 
It is this second dispersive interaction which can become problematic during the vacuum check and motivates us to engineer a more lossy bus mode.
\\
\\
When the transmon is being used for the vacuum check measurement and state tomography, we do not want it to interact with any photons in the bus mode via this dispersive interaction. The easiest way to ensure this is to wait for all photons in the bus to decay before proceeding with the vacuum check and state tomography. Emptying any photons in the bus mode as fast as possible minimizes the time we must wait before performing the vacuum check. This minimizes other sources of error such as cavity photon loss and self-Kerr which reduce entanglement fidelity. To mitigate these effects we increase the loss of the bus mode via a 3D Purcell filter. (A 2D on-chip Purcell filter would have also sufficed, although this would have requried fabrication). The internal loss rates of cable modes are typically $\sim$~\SI{100}{kHz}. With the Purcell filter, we can increase this to \SI{600}{kHz} allowing us to reach critical damping whilst preserving the coherence times of the transmons. Engineering a higher loss bus mode via a Purcell filter solves the challenges introduced by using the same transmon for cavity control and parametric coupling. Increasing the bus loss also allows us to demonstrate the efficacy of our scheme in a physical regime where previous approaches to deterministic state transfer would not be viable. 
\\
\\
Why do we opt for using a Purcell filter to increase the loss in the bus instead of using a more lossy cable such as one made from copper? Using a Purcell filter allows us to increase the loss of the bus mode without compromising the coherence times of the transmons. If we were to use a copper cable, this would likely Purcell-limit the transmons and would be detrimental for cavity state preparation and tomography. This limitation could be overcome if one were to instead use a separate coupling transmon and tomography transmon.
\\
\\
A description of the 3D purcell filter for the bus mode is shown in Fig.\,\ref{fig:purcell}. When measured at room temperature with a VNA, the Quality factor of the Purcell filter mode was measured to be $Q=170\pm10$ with a frequency of $\omega_\text{filter}= $\SI{5688}{MHz}, \SI{30}{MHz} above the bus mode frequency (although we cannot measure the Purcell mode frequency at cryogenic temperatures).
\\
\\
Ultimately, the need for a Purcell filter on the bus mode is not a strict requirement for the DMM scheme, but rather a consequence of using the same transmon to perform so many different tasks, which was done out of experimental convenience. 
\section{Parity measurements via frequency combs}
\label{app:freq_combs}
\begin{figure*}[h]
\includegraphics[scale = 0.73]{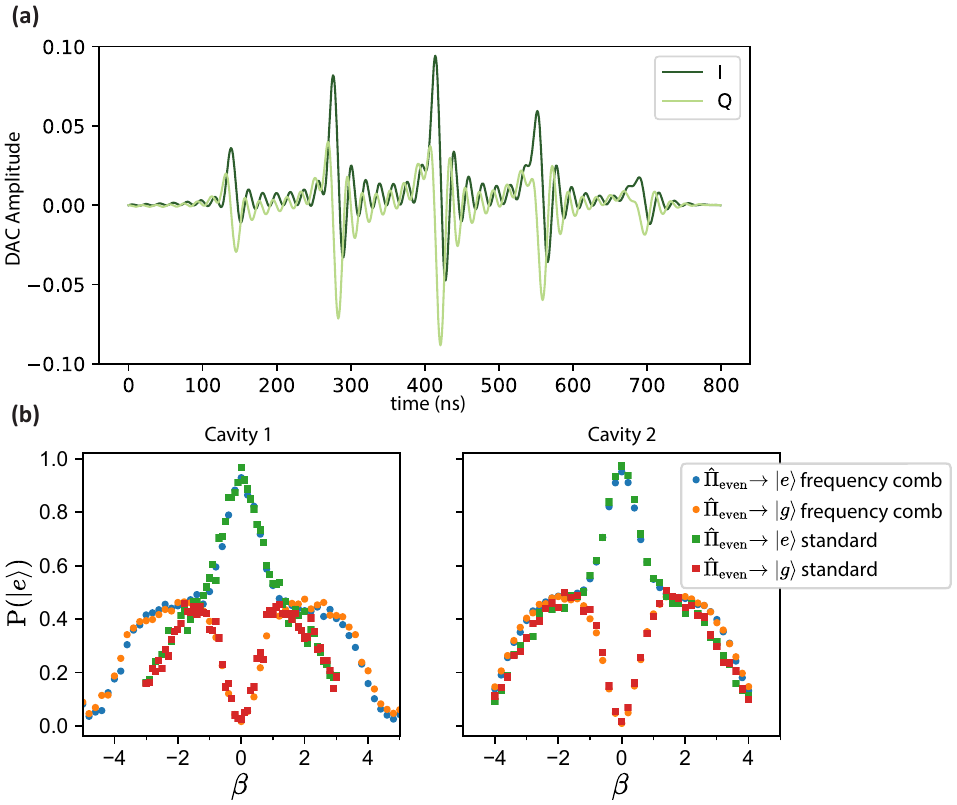}
\caption{\textbf{Improving parity measurements with combs of selective $\pi$-pulses} a) example pulse constructed from a comb of 8 selective Gaussian $\pi$-pulses with $\sigma=$\SI{200}{\nano\second} and $4\sigma$ duration. Each Gaussian is centered around frequency $\omega_n = \chi n +\chi'(n^2-n)$ for $n=\{0,2,4,6,8,10,12,14\}$. b) Comparing frequency combs to the standard parity pulse sequence. We prepare each cavity in vacuum and take a slice along the real axis of the Wigner function. For a parity measurement that works well for larger photon numbers, we expect $\text{P}(\ket{e})=0.5$ in the limit of large $\beta$. Mainly due to finite pulse bandwidth, we observe a drop off towards $\text{P}(\ket{e})=0$ as $\beta$ increases, regardless of which way we are mapping the parity to the transmon state. We see for cavity 1 that using a frequency comb instead results in a (steeper) drop-off that appears at almost twice the value of $\beta$ (around $\beta=3$ rather than $\beta=1.5$). For cavity 2, we observe similar performance for both the frequency comb and the standard parity sequence. Frequency combs are used to measure all Wigner functions. 
}
\label{fig:parity_comb}
\end{figure*}

 A usual parity measurement consists of a $\pi/2$ pulse on the transmon, a wait time $\pi/\chi_{at}$, another $\pi/2$ pulse about the same axis followed by transmon readout. If our transmon pulses are instantaneous, and if the interaction between the transmon and cavity takes exactly the form of a cross-Kerr interaction, then this sequence performs a QND photon number parity measurement. We will read out $\ket{e}$ if there are an even number of photons in the cavity whilst $\ket{g}$ signifies an odd number of photons. This pulse sequence was used to measure parity in \cite{Vlastakis2013}. When measuring the Wigner function, we take another data set where the amplitude of the second $\pi/2$ pulse is inverted, reversing the mapping of $\ket{g}$ and $\ket{e}$. The difference of these two data sets gives us the `symmetrized' Wigner function, which partially compensates for the following imperfections we now discuss.\\
\\
If we were to increase the number of photons in the cavities, eventually the $\pi/2$ pulses would no longer rotate the transmon, since the cross Kerr interaction would mean the $\pi/2$ pulses (which have finite-bandwidth) will always be off-resonant with the transmon. This biases our readout towards $\ket{g}$ at large cavity displacements and large numbers of photons in our cavities. This effect is partially corrected by symmetrizing the Wigner function. We no see  `Vignetting' in the Wigner function but still lose contrast at larger displacements. Other ways to combat this effect are to use larger bandwidth (shorter duration) $\pi/2$ pulses and to center the frequency of the $\pi/2$ pulse around the expected $\bar{n}$ in the cavity (although this does not account for the increase in variance in cavity photon number). Another detrimental effect is the so-called $\chi'$ correction to the dispersive interaction, which takes the form $H_{\chi'}/\hbar =\chi'\hat{a}^\dagger\hat{a}^\dagger\hat{a}\hat{a}\ket{e}\bra{e}$. Under the dispersive interaction, the frequency of the transmon shifts by $\chi n$ for $n$ photons in the cavity. With the $\chi'$ interaction included, this shift is now $\chi n +\chi'(n^2-n)$. This unwanted interaction also reduces the contrast of our parity measurement at large $\bar{n}$\\
\\
For cavity 2, $\chi_{at}/2\pi= $\SI{2}{MHz} and the effects discussed above have minimal effect on Wigner function measurements. For cavity 1, $\chi_{at}/2\pi= $\SI{3.7}{MHz}, almost twice as large and the effects discussed above become noticeable for the typical $\bar{n}$ of the cavity states we are measuring. 
\\
\\
We address all these problems in cavity 1 by replacing the `standard' pulse sequence for parity measurements with a comb of selective $\pi$-pulses. Although this pulse is longer in duration than the standard parity measurement pulse sequence, which leads to a slight overall reduction in the Wigner function contrast, this allows us to reliably measure the Wigner function of the cavity 1 state out to larger values of $\beta$.\\
\\
The comb is constructed by playing simultaneous selective $\pi$-pulses at frequencies $\omega_n = \omega_t + \chi n +\chi'(n^2-n)$ for even $n$ up to a maximum of $n=14$. An example of what this pulse looks like is shown in Fig.\,\ref{fig:parity_comb}a. To symmetrize the Wigner function, we use another frequency comb, this time containing frequencies for odd $n$ up to a maximum of $n=15$. Similar pulses have been used in other contexts \cite{ni2023beating}. We benchmark our pulse by taking a radial slice of the Wigner function of the vacuum state and comparing with the standard pulse sequence in Fig.\,\ref{fig:parity_comb}b and observe that for module 1, this approach is necessary to measure the Wigner function reliably at $\beta>1.5$

\newpage
\section{Compensating for the effects of Kerr}
\label{app:Kerr_comp}
\noindent Whilst the $\Hc$ Hamiltonian is being applied, and during the vacuum check measurement, the dark state in the cavity evolves under the local Kerr Hamiltonian, 
\begin{equation}
\mathcal{H}_{\text{Kerr}}/\hbar = \frac{K_1}{2}\hat{a}_1^\dagger\hat{a}_1^\dagger\hat{a}_1\hat{a}_1 + \frac{K_2}{2}\hat{a}_2^\dagger\hat{a}_2^\dagger\hat{a}_2\hat{a}_2    
\end{equation}
This evolution results in dark states no longer remaining exact dark states of the Hamiltonian $\Hc$, although the error induced on the entanglement fidelity is calculated to be small. (See App. \ref{app:budget}.) The other effect is that our final target entangled state is now given by the state
\begin{equation}
\ket{\psi_{\text{final}}}=\hat{\mathcal{U}}_{\text{Kerr}}\ket{\text{dark}'},    
\end{equation}
where 
\begin{equation}
\hat{\mathcal{U}}_{\text{Kerr}}=\hat{\mathcal{U}}_{\text{Kerr,1}}\hat{\mathcal{U}}_{\text{Kerr,2}}=e^{-\frac{i\theta_{K,1}}{2}(\hat{a}_1^\dagger\hat{a}_1^\dagger\hat{a}_1\hat{a}_1)}e^{-\frac{i\theta_{K,2}}{2}(\hat{a}_2^\dagger\hat{a}_2^\dagger\hat{a}_2\hat{a}_2)}.    
\end{equation}
The angles $\theta_{K,1}$ and $\theta_{K,2}$ are measured directly for each entanglement experiment. This is achieved by preparing the cavities in the dark state $\ket{\alpha}_1\ket{-\alpha}_2$, enacting $\Hc$ and the vacuum check measurements and then measuring the single mode Wigner function on each cavity. With maximum likelihood estimation, we reconstruct the density matrix from the measured Wigner function, and find the angle, $\theta_{K,i}$ which maximizes the overlap between this density matrix and the state $e^{-\frac{i\theta_{K,i}}{2}(\hat{a}^\dagger\hat{a}^\dagger\hat{a}\hat{a})}\hat{\Pi}_{\bar{0}}\ket{\alpha}$, where $\hat{\Pi}_{\bar{0}} = \mathds{1}-\ket{0}\bra{0}$
\\
\\
The $\theta_{K,i}$ values we measure are slightly lower from what we would expect from the unpumped cavity self-Kerrs acting for $\approx$\SI{3}{\micro\second}, because the pumps that enact $\Hc$ are also observed to reduce the magnitude of the cavity self-Kerrs. Once these values are known, on system 2, we alter the state transfer conditions for the OCT pulses that decode the cavity state onto the qubit, by making sure to apply the Kerr unitary to the cavity states we wish to decode when optimizing the OCT pulse. 
\\
\\
For the highest reported teleportation fidelity, we do not compensate for the Kerr in cavity 1, but instead redefine our logical codewords to include this Kerr effect. We see this as visible distortions in the Wigner functions in Fig. \ref{fig:ent_tele_full}, and redefine the logical codewords of cavity 1 to compensate. For the teleportation Wigner functions presented in the main text, we actively correct for Kerr in cavity 1, to make the required Pauli corrections from teleportation more evident visually.
\\
\\
For the entanglement experiment in \ref{fig:Fig3}, we apply a blue detuned pump \cite{YaxingKerrCancellation} during the vacuum check, with a detuning of \SI{+60}{MHz} from the $\omega_{ge}$ transition of transmon 1. The stark shift on transmon 1 from this pump is measured to be \SI{-11.4}{MHz}, and we must adjust the  frequency of the selective $\pi$-pulse on transmon 1 to account for this. We measure an increase in the Bell state fidelity by 1\% when we apply this pump, although this may be related to maximum-likelihood reconstruction performing worse for states that have undergone significant Kerr evolution. We only investigated the effects of applying the Kerr cancellation pump at $\alpha=1.414$
\\
\\
The other way we compensate for Kerr is used in the teleportation data in \ref{fig:Fig4} where we replace the selective $\pi$-pulse with a SNAP-like pulse that excites transmon 1 if there are zero photons in the cavity, and otherwise performs a SNAP gate that undoes the Kerr evolution and leaves transmon 1 in the ground state. We can write the action of this pulse (assuming transmon 1 begins in $\ket{g}$) as
\begin{equation}
    \ket{e}\bra{g}\otimes\ket{0}\bra{0}_1+\ket{g}\bra{g}\otimes\hat{\mathcal{U}}^\dagger_{\text{Kerr,1}}\hat{\Pi}_{\bar{0}}. 
\end{equation}
Such a pulse is readily found with OCT and resembles a frequency comb. This time, we find this pulse reduces the teleportation fidelity by 1\%, with additional error expected from transmon decoherence when performing the SNAP unitary, which requires the transmon to be excited out of its ground state. 
\newpage 
\section{System Parameters}
\label{app:params}

\begin{table}[h]
\begin{tabular}{l|c|c}
\toprule
\textbf{Mode}  & \multicolumn{1}{l|}{\textbf{Parameter}} & \multicolumn{1}{c}{\textbf{Frequency (MHz)}} \\ \hline
Cavity 1            & $\omega_{a_1}/2\pi$                               & 6509                           \\
Cavity 2            & $\omega_{a_2}/2\pi$                              & 6487                           \\
Bus mode            & $\omega_{b}/2\pi$                               & 5658                           \\
Bus Purcell filter  & $\omega_{\text{filter}}/2\pi$                            & 5688                            \\
Transmon 1          & $\omega_{t_1}/2\pi$                               & 5358                           \\
Transmon 2          & $\omega_{t_2}/2\pi$                               & 4759                           \\
Pump $x_1$             & $\omega_{x_1}/2\pi$                               & 5425.0                         \\
Pump $y_1$             & $\omega_{y_1}/2\pi$                               & 6271.76                        \\
Pump $x_2$             & $\omega_{x_2}/2\pi$                                & 4950.381                       \\
Pump $y_2$             & $\omega_{y_2}/2\pi$                                & 5775.11                       \\
Readout resonator 1 & $\omega_{\text{RO1}}$                              & 7981                           \\
Readout resonator 2 & $\omega_{\text{RO2}}$                              & 7895 \\          \toprule               
\end{tabular}
\caption{Measured mode and pump frequencies used in the experiment}
\end{table}

\begin{table}[h]
\begin{tabular}{l|c|c}
\toprule
\textbf{Coupling}                  & \textbf{Parameter} & \textbf{Frequency (MHz)} \\ \hline
Cavity 1 - Transmon 1 dispersive   & $\chi_{a_1t_1}$          & -3.75                     \\
Cavity 2 - Transmon 1 dispersive   & $\chi_{a_2t_2}$           & -2.2                      \\
Bus mode - Transmon 1 dispersive   & $\chi_{bt_1}$           & -2.1                      \\
Bus mode - Transmon 2 dispersive   & $\chi_{bt_2}$           & -2.5                      \\
Cavity 1 self-Kerr                 & $K_1$                 & -0.023                    \\
Cavity 2 self-Kerr                 & $K_2$                  & -0.007                    \\
Cavity 1 - Transmon 1 $\chi'$   & $\chi'_{a_1t_1}$          & +0.017                     \\
Cavity 2 - Transmon 2 $\chi'$   & $\chi'_{a_2t_2}$          & +0.004                    \\
Transmon 1 anharmonicty            & $\alpha_{t1}$           & -182                      \\
Transmon 2 anharmonicty            & $\alpha_{t2}$           & -187                      \\
Transmon 1 - Transmon 2 dispersive & $\chi_{t_1t_2}$          & 0.05     \\               
\toprule
\end{tabular}
\caption{Measured static mode couplings present in the experimental hardware.}
\end{table}

\begin{table}[h]
\begin{tabular}{l|c|c|c}
\toprule
\textbf{Mode}       & $\mathbf{T_1}$ or $\mathbf{T_2\,(\mu s)}$    & $\mathbf{\kappa}$\,\textbf{(kHz)} & \textbf{Quality Factor} \\
\hline
Bus mode            & 0.27               & $600\pm10$            & 9400                    \\
Purcell filter mode &                    &                    & $170\pm10$                 \\
Cavity 1 ($T_1$)           & $385\pm5$      &                    &                         \\
Cavity 1 ($T_2^R$)           & $235\pm10$      &                    &                         \\
Cavity 2 ($T_1$)           & $520\pm15$      &                    &                         \\
Cavity 2  ($T_2^R$)          & $240\pm10$      &                    &                        \\
Transmon 1  ($T_1$)        & $31\pm4$ &                    &                         \\
Transmon 1  ($T_2^R$)        & $27\pm1$ &                    &                         \\
Transmon 1  ($T_2^E$)        & $34\pm1$ &                    &                         \\
Transmon 2 ($T_1$)         & $52\pm1$ &                    &                         \\
Transmon 2  ($T_2^R$)        & $21\pm1$ &                    &                         \\
Transmon 2  ($T_2^E$)        & $36\pm1$ &                    &                         \\
Readout resonator 1 &                    & $360\pm5$                &                         \\
Readout resonator 2 &                    & $730\pm5$                &      \\             \toprule     
\end{tabular}
\caption{Measured coherence times of modes used in the experiment.}
\end{table}

\newpage
\section{Wiring diagram}
Here we show the fridge wiring diagram used in the experiment, including the readout chain, drive line filtering and the way we source the four pump tones needed to enact $\Hc$.
\label{app:wiring}
\begin{figure*}[h] 
\includegraphics[scale = 0.84]{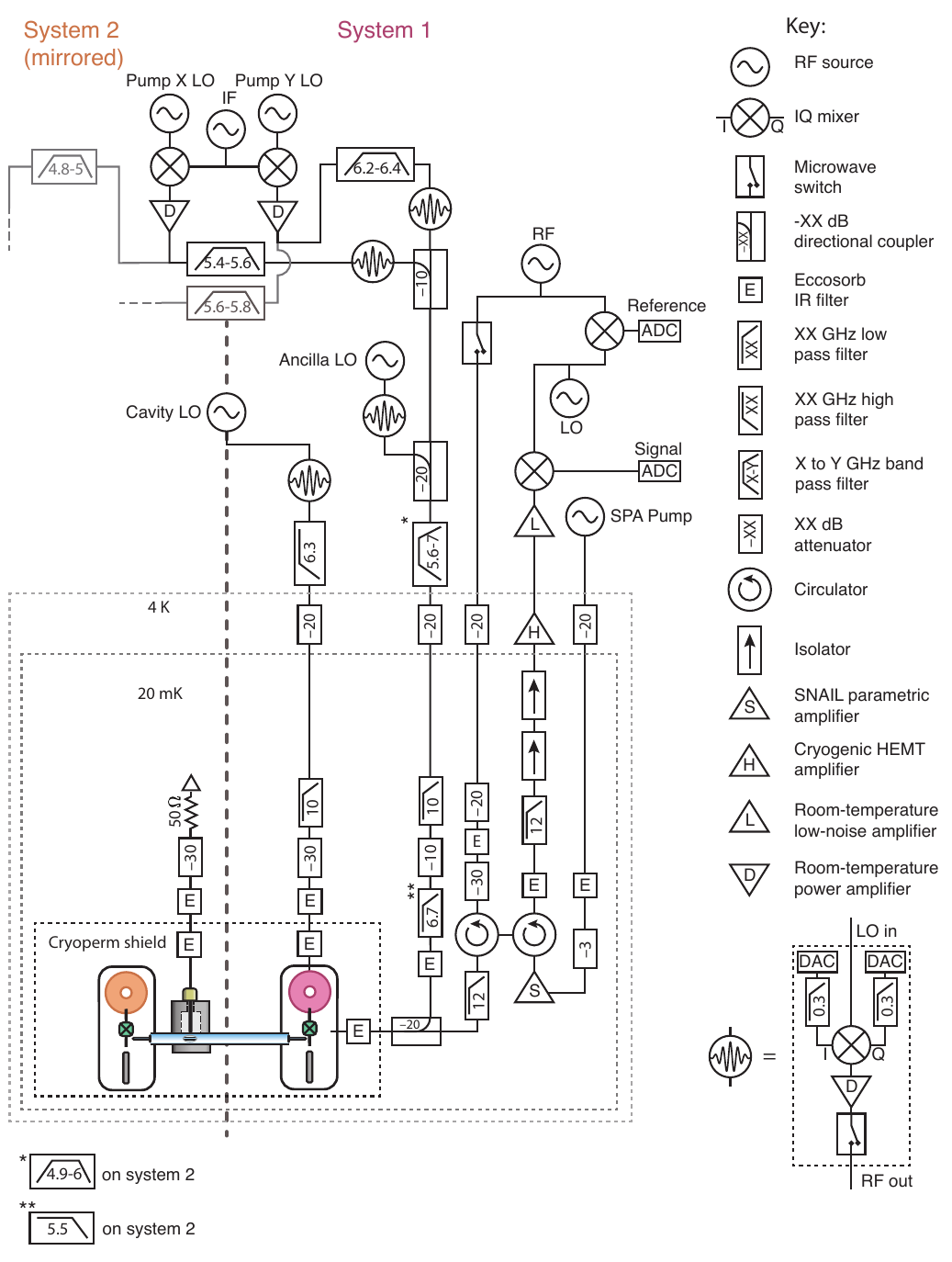}
\caption{\textbf{Fridge wiring diagram}. Except where noted, the input/output lines on system 2 are exact duplicates of the lines shown for system 1.}
\label{fig:wiring}
\end{figure*}
\section{Pump calibration, sourcing and phase locking}
\label{app:pump_cal}
\begin{figure*}[h] 
\includegraphics[scale = 1.0]{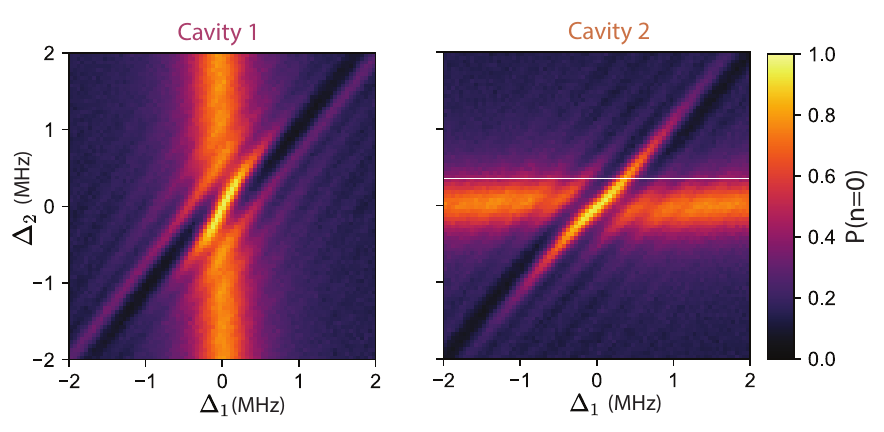}
\caption{\textbf{Pump calibration to account for stark shift}. We sweep the difference in the pump frequencies, $\Delta_i=\omega_{y_i}-\omega_{x_i}$ by sweeping the value of $\omega_{y_i}$. We choose to sweep $\omega_{y_i}$ because this pump is far detuned from the transmon and so this does not drastically change the normalized pump amplitude. To perform this calibration we prepare the coherent state $\ket{\alpha}_1\ket{\alpha}_2$ (here $\alpha = \sqrt{2}$), apply all four pump tones for \SI{2}{\micro\second}, and then measure the cavity states with selective $\pi$-pulses to see if it is in the vaccuum state. When $\Delta_1$ is close to zero, but $\Delta_2$ is far from zero, we see a vertical bar showing the $g_1$ process on resonance, but the $g_2$ process off-resonance. The same thing happens on the right plot but with $\Delta_1$ and $\Delta_2$ reversed, which gives a horizontal bar. The ideal pump detunings are found from where the vertical and horizontal bar intersect. As we approach $\Delta_1=\Delta_2=0$, we see the appearance of fringes, a result of the cavity dynamics now becoming dependent on the relative phases between both cavities. This signature also verifies our pumps are phase-locked correctly.}
\label{fig:pump_cal}
\end{figure*}

\noindent A total of four pump tones must be sourced to enact the Hamiltonian $\Hc$. When activating a beamsplitter process (Eq. \ref{eq:bs}) via four-wave mixing, we must satisfy the frequency matching condition $|\omega_{a_i}-\omega_b|=|\omega_{y,i}-\omega_{x,i}|$. We find the engineered beamsplitter is the `cleanest' (lowest transmon heating, few spurious parametric resonances) when one of the pumps, pump $x$ is placed around \SIrange{100}{200}{MHz} above the transmon $\omega_{ge}$ frequency and the other pump, pump $y$ is placed around \SIrange{900}{1000}{MHz} above. In system 1, pump $x_1$ is designated the `strong pump', since it induces the most stark shift on transmon 1. In system 2, both pumps induce similar amounts of start shift on transmon 2.
\\
\\
When we actuate both beamsplitters simultaneously, the frequency matching conditions change for both beamsplitter processes, since every pump tone induces a stark shift on the bus mode. We now walk through how we tune up the pump amplitudes and detunings to enact $\Hc$.
\\
\\
The first step is to separately tune up the $g_1$ and $g_2$ beamsplitter processes and to match their beamsplitter rates as closely as possible. This is achieved by preparing a coherent state, applying the beamsplitter for a variable time and then measuring the probability the cavity is in the vacuum state via selective $\pi$-pulses on the transmon. By fitting the time dynamics, we can tune both $g_1$ and $g_2$ to be close to \SI{160}{kHz}, although since we are in the slightly overdamped regime, the lack of oscillations gives an uncertainty of $\sim$~\SI{5}{kHz}.
These pump conditions are also used to Q-switch cavity states to the bus mode to reset them to vacuum at the end of an experimental shot, by first Q-switching cavity 1 to the bus mode for \SI{4}{\micro\second}, waiting \SI{5}{\micro\second}, and then Q-switching cavity 2 to the bus mode for another \SI{4}{\micro\second}. 
\\
\\
After $g_1$ and $g_2$ have been tuned up individually, we must tune them up simultaneously. This mostly amounts to sweeping the pump $y$ frequencies, $\omega_{y_1}$ and $\omega_{y_2}$ until both processes are on resonance when activated simultaneously. This callibration procedure is shown in Fig.\,\ref{fig:pump_cal}.
\\
\\
Next we verify these pump frequencies are close to optimum by looking at time evolution vs. relative cavity phase (as shown in Fig.\,\ref{fig:Fig2}) and looking for any asymmetric or skewed dynamics. The final step of the calibration is to run the full entanglement protocol and measure the joint photon number parity of the final dark state. The joint photon number parity should be as `odd' as possible and is a good proxy for entanglement fidelity. We fine tune the pump frequencies $\omega_{y_1}$ and $\omega_{y_2}$ (sensitive to \SI{10}{kHz} changes), as well as the pump $y_2$ DAC amplitude (sensitive to 2\% changes) to minimize the joint-parity.  
\\
\\
We must repeat this pump tuneup procedure whenever we change the value of $\alpha$ we use for our entangled state generation. Typically, these adjustments are of the order \SI{10}{kHz} and are thought to result from cavity self-Kerr, meaning that the resonance condition is weakly dependent on the number of photons in the cavities. 
\\
\\
The four pump tones must be appropriately phase-locked with each other, to ensure we enact the $\Hc$ Hamiltonian on each shot with the correct relative phase between the $g_1$ and $g_2$ beamsplitter tones. Each tone is synthesized by sending an LO tone through an IQ mixer where it is combined with a single-sideband (SSB) tone from the FPGA controller. All SSBs remain phase-locked with each other over the course of a single shot, since they are sourced from the same FPGA controller, but drift apart on the timescale of seconds, and so their frequencies are reset after every experimental shot. Note that the two cavity drives at $\omega_{a_1}$ and $\omega_{a_2}$ must be phase locked with each other (and hence share the same LO generator) but do not need to be phase locked with any of the pumps that actuate $\Hc$.
\\
\\
We enact $\Hc$ with fixed relative phase when pump $x_1$ is phase-locked with pump $x_2$, and pump $y_1$ is phase locked with pump $y_2$; the same phase-locking condition is required in\,\cite{Luke2020}. This is typically achieved by sourcing both pump $x_1$ and $x_2$ LO tones from the same signal generator, and using SSB tones to synthesize the final pump tones. Achieving this phase-locking is more involved than merely sourcing pump tones from the same LO. 
The $\omega_{ge}$ transitions of transmon 1 and transmon 2 are detuned by $\sim$~\SI{500}{MHz}. This is by design, since we can have an appreciable $ZZ$ coupling between the transmons mediated by the bus mode if they are too close in frequency. With this detuning, we observe a $ZZ$ cross talk of \SI{50}{kHz}. For phase locking, this means we must phase lock pump tones that differ in frequency by $\sim$~\SI{500}{MHz} (in order to place each pump at the desired frequency above the transmons.) However, our FPGA controller can only synthesize SSB tones in the range $\pm$\SI{125}{MHz}.
\\
\\
As shown in Fig.\,\ref{fig:wiring}, we solve this problem by using another signal generator, which we call the `IF' generator. We set its frequency to \SI{300}{MHz}, whilst the pump X LO generator is set to \SI{5200}{MHz} and the pump Y LO generator set to \SI{6000}{MHz}. With the use of additional mixers, we can generate phase-locked tones at $5200\pm300$\,MHz and $6000\pm300$\,MHz. By using four tunable narrow-band cavity filters, we can pick off just one of these tones to source the IQ mixers. The pump $x_1$ IQ mixer LO is at \SI{5500}{MHz}, the pump $x_2$ IQ mixer LO is at \SI{4900}{MHz}, the pump $y_1$ IQ mixer LO is at \SI{6300}{MHz} and the pump $y_2$ IQ mixer LO is at \SI{5700}{MHz}.
\section{Measuring loss of the bus mode}
\begin{figure*}[h] 
\includegraphics[scale = 0.4]{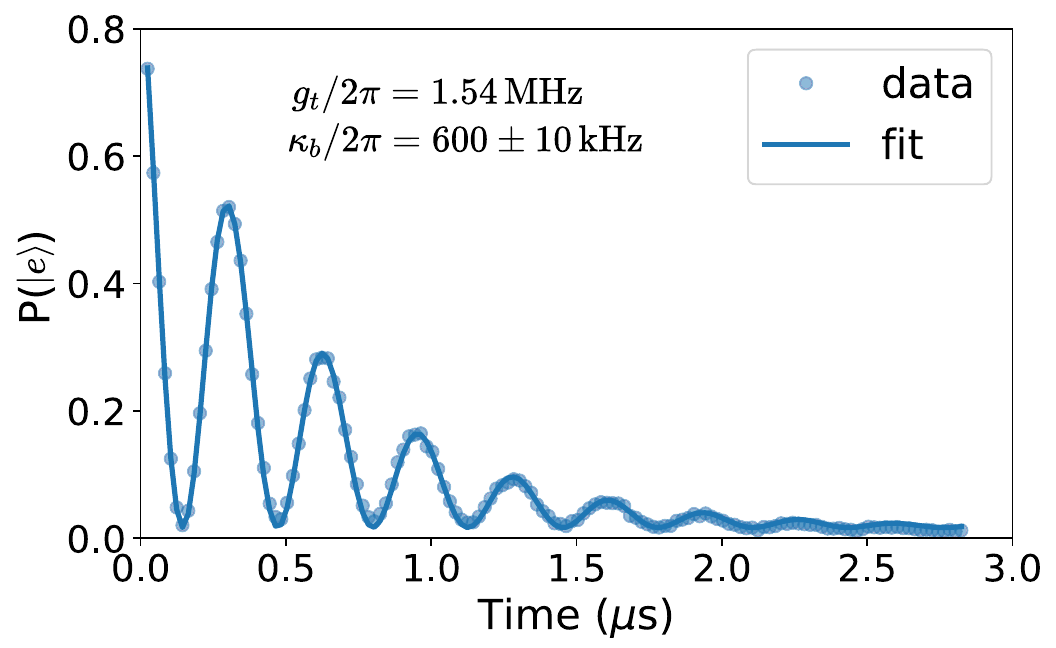}
\caption{\textbf{Measuring the bus loss, }$\mathbf{\kappa_b}$ We prepare transmon 2 in $\ket{e}$, and actuate a sideband transition with the bus mode. The value of $\kappa_b$, as well as the sideband coupling rate, $g_t$ are both extracted from the fit.}
\label{fig:kappa_b_fit}
\end{figure*}

To measure $\kappa_b$, the loss of the bus mode it is desirable to engineer $g>\kappa_b$ such that excitations oscillate multiple times in and out of the bus. We cannot engineer such a large $g$ with our cavity-bus coupling but we can engineer a sideband coupling between the transmon and the bus mode, which is another four-wave mixing process that is actuated if we satisfy the pump condition $|\omega_{t_i}-\omega_b|=|\omega_{y,i}-\omega_{x,i}|$. Since the transmon mode has a much higher participation in the Josephson junction than the cavity mode, we can engineer a much larger coupling, ($\times10$ faster) without needing to increase the pump power. We choose to use transmon 2, which has the higher $T_1$ and measure $\kappa_b$ by first preparing the transmon in the excited state, actuating the coupling 
\begin{equation}
\mathcal{H}_t/\hbar = g_t\left(\ket{g}\bra{e}\hat{b}^\dagger+\ket{e}\bra{g}\hat{b}\right).     
\end{equation}
After a variable time, we measure the state of transmon 2. Since $g_t>\kappa_b$, we observe many oscillations. It is important that the transmon $T_1$ is long-lived relative to the bus mode lifetime, such that decay in the bus mode dominates the energy decay rate. We fit the data shown in Fig.\,\ref{fig:kappa_b_fit} to the analytical solution $|a(t)|^2$ to the set of quantum Langevin equations that describe this coupling and dissipation:

\begin{align}
\label{eq:transmonbus}
\begin{split}
    \dot{a}(t) & = i \, g_t\:b(t)-\frac{\kappa_a}{2}\, a(t)\\
    \dot{b}(t) & = -i \, g_t\:a(t)-\frac{\kappa_b}{2}\, b(t),
\end{split}
\end{align}
with the initial conditions $a(0) = 1, b(0)=0$. We set $\kappa_a = 1/T_1$, assuming that the $T_1$ of transmon 2 does not significantly change when the coupling is actuated, which is why we quote a $\pm$\SI{10}{kHz} uncertainty on $\kappa_b$. 

\section{Error budget and the optimum value of $\alpha$}
\label{app:budget}
\noindent The highest fidelity entanglement is measured at $\alpha = \sqrt{2}$ with an infidelity of $8\pm1\%$, not correcting for any SPAM errors. We identify three dominant error sources that contribute to this infidelity. The largest contribution is from photon loss in the cavity modes, a local error. We can estimate this from the time of the protocol (from the initial cavity displacements until state tomography). This time is \SI{5.6}{\micro\second} (800 ns OCT SNAP pulse + 2092 ns $\Hc$ pumps + 1200 ns selective $\pi$-pulses + 1500ns transmon readout = 5592 ns = $t_{\text{protocol}}$). During this time, a single photon loss event in either cavity would take us to the state $(\ket{\alpha}\ket{-\alpha}+\ket{-\alpha}\ket{\alpha})/\sqrt{2}$, which orthogonal to our desired entangled state. The probability of this occurring is given by
\begin{equation}
    p_{\text{loss}}=|\alpha|^2 t_{\text{protocol}} \left(\frac{1}{T_{1,a_1}} + \frac{1}{T_{1,a_2}}\right)\approx5\%,
\end{equation}
if we assume the cavity lifetimes remain unchanged throughout the protocol.  
\\
\\
The next dominant source of error is a SPAM-type error that comes from our OCT decode pulse used for mapping the logical state from cavity 2 onto transmon 2. We estimate this error to be $p_{\text{decode}} = 1.7\%$ from the encode/decode calibration described in App. \ref{app:twobit_tomo}
\\
\\
The last error we suspect is the `false positive' rate of the vacuum check. This is defined to be
\begin{equation}
    \Fdmm = \frac{p(gg|\text{bright})}{p(gg|\text{dark})+p(gg|\text{bright})}
\end{equation}
We can measure the numerator and denominator directly from Fig.\,\ref{fig:Fig2} when we take a linecut at \SI{2}{\micro\second}. We measure $p(gg|\text{bright}) = 1.5\%$ and $p(gg|\text{dark}) = 70\%$, close to the theoretical value of $1-2e^{-|\alpha|^2}+e^{-2|\alpha|^2}\approx73\%$ from the overlap with vacuum. The 3\% difference is due to transmons being excited out of their ground states due to rethermalization and additional heating induced by the pump tones.  This gives $\Fdmm=2.1\%$.\\
\\
The value of $p(gg|\text{bright})$ is larger than what we would expect from individual vacuum checks. Experimentally, we measure $p(g_{1(2)}|\text{bright}) = 7\% (5\%)$ when the vacuum check is performed via transmon 1 (2). If these errors were due solely to uncorrelated errors as one may expect from simple transmon decoherence, we would expect $p(gg|\text{bright})$ to be around four times lower than its experimentally measured value. Although we were unable to determine the source of this `correlated error' in our vacuum check measurements, imperfections in the vacuum check only make up a small fraction of our total error budget.
\\
\\
Our total error budget for Bell state generation can be written as
\begin{equation}
p_{\text{loss}} + p_{\text{decode}}+ \Fdmm = 9\%,   
\end{equation}
which is in fairly good agreement with our measured fidelity at $\alpha = 1.414$.
\\
\\
This error model also informs us why we should expect an optimum value of $\alpha$ that maximizes the entanglement fidelity. At larger values of $\alpha$, $p_{\text{false positive}}$ saturates to 1\% whereas $p_{\text{loss}}$ continues to increase quadratically. At smaller values of $\alpha$, the success probability, set by $p(gg|\text{dark})$, drops rapidly, causing $p_{\text{false positive}}$ to increase rapidly as well. \\
\\
Our simple error model predicts an optimum $\alpha$ of 1.09 whilst experimentally we find entanglement fidelity is maximized near $\alpha = \sqrt{2}$. We observe that entangled states with $\alpha<\sqrt{2}$ have lower fidelities than expected. From the reconstructed density matrix of cavity 1, we observe an excess of leakage to the vacuum state $\ket{0}$ (despite this state supposedly passing the vacuum check), also indicated by the decrease in the $II$ Pauli bar, which is not predicted by this error model.  

\section{Joint Wigner Tomography}
\begin{figure*}[h] 
\includegraphics[scale = 0.7]{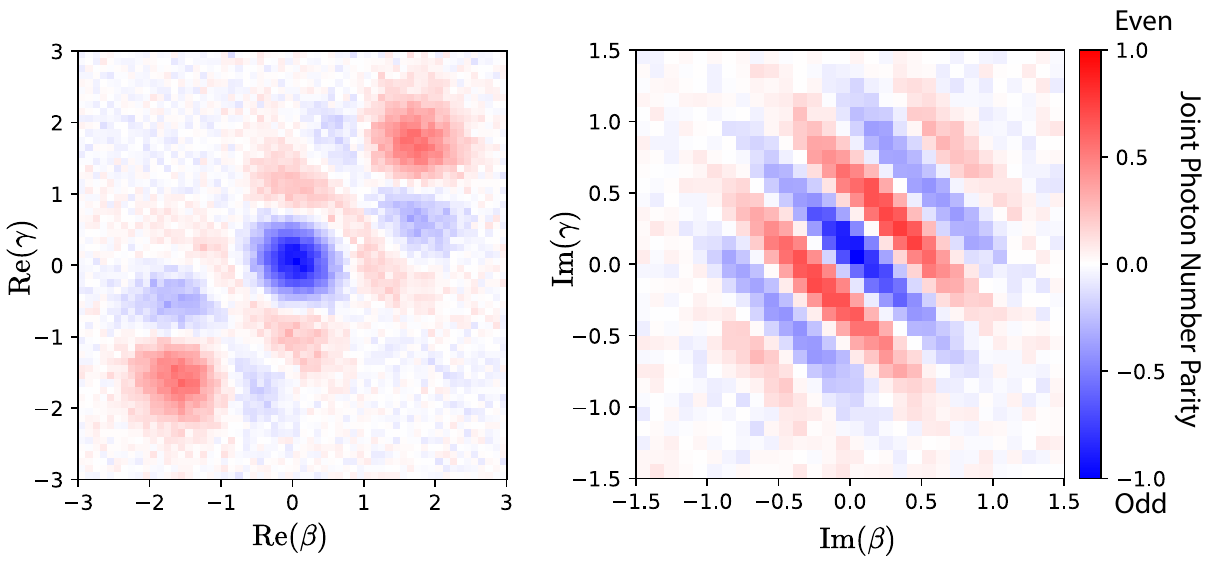}
\caption{\textbf{2D Slices of the joint Wigner function for the entangled state generated at $\alpha_0 = 1.6$} On the left we fix $\beta$ and $\gamma$ displacements to be real-valued. The principal features of this plot are correlations visible around $(\beta,\gamma) = (1.6,-1.6)$ and $(\beta,\gamma) = (-1.6,1.6)$, which verifies we are confined to the $\{\ket{\pm\alpha,\mp\alpha}\}$ manifold. The negative joint parity centered around the origin verifies we are in an entangled state. The fainter features are expected from removing states containing zero photons from the cat-in-two-boxes state. On the right we fix $\beta$ and $\gamma$ to be imaginary, and see interference fringes, another signature of entanglement for this state.}
\label{fig:jwig}
\end{figure*}

\noindent Another way to verify we have entanglement in our final cavity state is to perform joint winger tomography. Although harder entanglement fidelity is harder to quantify in this way, due to the 4D nature of the joint Wigner function, taking 2D slices of the joint Wigner function nonetheless gives a striking visualization of the entanglement in the form of non-local interference fringes and has been presented in several previous works where the cat-in-two-boxes state was the state of interest. 
\\
\\
The joint Wigner function is given by the displaced joint-parity operator. For a pure state, this is given by
\begin{equation}
    W_J(\beta,\gamma) = \braket{\psi_{12}|\hat{D}_{\beta,1}^\dagger \hat{D}_{\gamma,2}^\dagger\hat{P}_1\hat{P}_2\hat{D}_{\beta,1}\hat{D}_{\gamma,2}|\psi_{12}}.
\end{equation}
We perform joint Wigner tomography experimentally by performing the two displacements, $D_{\beta,1}D_{\gamma,2}$, and then measuring the individual photon number parities of each cavity. Multiplying these parity outcomes together on each experimental shot gives us the joint-parity (which ranges from -1 to 1). Averaging over many shots gives us the joint-Wigner function. In Fig.\,\ref{fig:jwig}. we plot the real-real and imaginary-imaginary slices of the joint Wigner function for the entangled state generated with $\alpha=1.6$
\newpage
To obtain these plots, we must appropriately symmetrize the Wigner function, just as we do for single mode Wigner. In the single mode case, we achieve this by taking one data set where the transmon is mapped to $\ket{e}$ which we call $W^{e}(\beta)$ if the parity is even, and another where the transmon is mapped to $\ket{g}$ which we call $W^{g}(\beta)$. The symmetrized Wigner is obtained by subtracting these (averaged) data sets i.e. 
\begin{equation}
    W(\beta) = \frac{W^{e}(\beta)-W^{g}(\beta)}{2}
\end{equation}
This technique can compensate for systematic errors such as the parity measurements becoming more biased to the $\ket{g}$ outcome at larger cavity displacements. When we measure the joint Wigner with local parity measurements, we can measure the Wigner function in four different ways: $W^{g_1,g_2}(\beta,\gamma),W^{g_1,e_2}(\beta,\gamma),W^{e_1,g_2}(\beta,\gamma)$ and $W^{e_1,e_2}(\beta,\gamma)$. The symmetrized joint Wigner is given by
\begin{equation}
    W_J(\beta,\gamma) = \frac{W_J^{g_1,g_2}(\beta,\gamma)-W_J^{e_1,e_2}(\beta,\gamma)-W_J^{e_1,g_2}(\beta,\gamma)+W_J^{g_1,e_2}(\beta,\gamma)}{4}
\end{equation}
and is used to process the data in Fig.\,\ref{fig:jwig}.
\section{Bell Measurements}
\label{app:Bell_msmts}
Here we describe how the Bell measurements were performed to obtain the outcomes $(m_1,m_2)$ for state teleportation. We use two successive measurements of transmon 2 to to obtain these values.\\
\\
To see how our circuit can perform a Bell measurement on entangled states between transmon 2 and cavity 2, it is helpful to consider what happens if cavity 2 is encoded in the cat code, instead of our modified basis. The desired Bell measurement amounts to performing a destructive measurement that distinguishes between the following four states:
\begin{align}
    \begin{split}
    \ket{\psi_1} &\propto \ket{\alpha}_2\ket{g}_2 + \ket{-\alpha}_2\ket{e}_2 \\  
    \ket{\psi_2} &\propto \ket{\alpha}_2\ket{g}_2 - \ket{-\alpha}_2\ket{e}_2 \\ 
    \ket{\psi_3} &\propto \ket{\alpha}_2\ket{e}_2 + \ket{-\alpha}_2\ket{g}_2 \\  
    \ket{\psi_4} &\propto \ket{\alpha}_2\ket{e}_2 - \ket{-\alpha}_2\ket{g}_2 \\ 
    \end{split}
\end{align}
The value of $m_1$ distinguishes between states $\{\ket{\psi_1},\ket{\psi_3}\}$ and $\{\ket{\psi_2},\ket{\psi_4}\}$ whereas the value of $m_2$ distinguishes between states $\{\ket{\psi_1},\ket{\psi_2}\}$ and $\{\ket{\psi_3},\ket{\psi_4}\}$. Once we know $(m_1,m_2)$, we know which state, $\ket{\psi_{i=1,2,3,4}}$ we started in, completing our Bell measurement. 
The CNOT gate is performed simply by waiting time $\pi/\chi_{a_2t_2}$. This causes the states to evolve as
\begin{align}
    \begin{split}
    \ket{\psi_1} &\propto \ket{\alpha}_2\ket{g}_2 + \ket{-\alpha}_2\ket{e}_2 \rightarrow \ket{\alpha}_2 (\ket{g}+\ket{e})_2 \\  
    \ket{\psi_2} &\propto \ket{\alpha}_2\ket{g}_2 - \ket{-\alpha}_2\ket{e}_2 \rightarrow \ket{\alpha}_2 (\ket{g}-\ket{e})_2\\ 
    \ket{\psi_3} &\propto \ket{\alpha}_2\ket{e}_2 + \ket{-\alpha}_2\ket{g}_2 \rightarrow \ket{-\alpha}_2 (\ket{g}+\ket{e})_2\\  
    \ket{\psi_4} &\propto \ket{\alpha}_2\ket{e}_2 - \ket{-\alpha}_2\ket{g}_2\rightarrow \ket{-\alpha}_2 (\ket{g}-\ket{e})_2 \\ 
    \end{split}
\end{align}
We can see that we measure $m_1$ by measuring the transmon in the $\ket{\pm}$ basis after this wait time. We then conditionally reset the transmon and measure cavity 2 in the $\ket{\pm \alpha}$ basis by using an OCT pulse to map these cavity states to the g/e states of the transmon before reading out the transmon again. This whole sequence only imparts around a 2\% additional error on the teleportation fidelity, compared to the Bell state fidelity. We use the modified cavity 2 logical basis to find the OCT pulse and account for effects such as self-Kerr. The pulse sequence is shown below.
\begin{figure*}[h] 
\includegraphics[scale = 1.0]{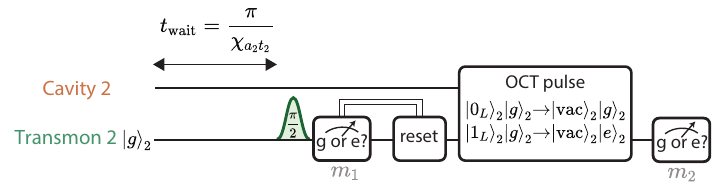}
\caption{\textbf{Pulse  sequence for realizing a Bell measurement between cavity 2 and transmon 2}. We extract two bits of information, ($m_1,m_2$) which constitutes a Bell measurement. First, we use the dispersive interaction and simply wait for time $t_\mathrm{wait}$to perform a CNOT gate between the transmon and cavity state. We then readout the transmon in its $X$ basis by performing a $\pi/2$ pulse followed by standard transmon dispersive readout followed by conditional transmon reset. To read out the state of the cavity, we perform an OCT pulse to transfer the logical state of the cavity onto the g-e manifold of the tranmon, followed by another transmon dispersive readout. This allows us to measure cavity 2 in its logical Z basis after the CNOT gate.}
\label{fig:OCTpulse}
\end{figure*}
\newpage
\section{Applicability to other  encodings}
\noindent The DMM scheme seems particularly well-suited to the two-legged cat code, due to the convenient simple interference between coherent states under $\Hc$. 
It is however possible to use a variation of the DMM scheme to produce Bell states in the dual-rail encoding, albeit with a reduced success probability of 1/8. The dual-rail logical codewords are defined to be $\ket{0_L}=\ket{01}$, $\ket{1_L} = \ket{10}$ and are encoded in a pair of harmonic oscillators. This scheme stems from the observation that if one prepares the initial state $\ket{1,0}_{12}$ and actuates $\Hc$ indefinitely then the system reaches a non-vacuum steady state given by
\begin{equation}
    \ket{\text{steady}}\bra{\text{steady}}_{12}=\frac{\ket{\Phi_-}\bra{\Phi_-}_{12}}{2} + \frac{\ket{0,0}\bra{0,0}_{1,2}}{2}
\end{equation}
where 
\begin{equation} 
\ket{\Phi_-}_{12} = \frac{\ket{1,0}_{12}-\ket{0,1}_{12}}{\sqrt{2}}
\end{equation}
We reach this rather unusual steady state because the intial state can be written as a superposition of the dark state, $\ket{\Phi_-}_{12}$ and the bright state $\ket{\Phi_+}_{12}=(\ket{1,0}_{12}+\ket{0,1}_{12})/\sqrt{2}$. This scheme is subtly different in the sense that we rely on bus loss to reach this particular steady state. 

The idea is to distill a perfect Bell state in the dual-rail encoding out of two copies of this steady state using local joint parity measurements in each module. 

If we repeat the protocol twice amongst four cavity modes, then 1/4 time, the cavity states can be described as
\begin{align}
\begin{split}
\ket{{\Phi_-}_{12}}\ket{{\Phi_-}_{34}} &= \frac{ \ket{1,0,1,0}_{1234}+\ket{0,1,0,1}_{1234}}{2}\\&+\frac{\ket{1,0,0,1}_{1234}+\ket{0,1,1,0}_{1234}}{2}
\end{split}
\end{align} 
If we arrange our cavities such that modes 1 and 3 are housed in one module, comprising one dual-rail qubit and modes 2 and 4 are housed in another module (for the other dual-rail qubit), then we can measure non-destructively the joint-photon number parity of each of these cavity pairs. Half of the time, we obtain odd joint-photon number parity outcomes for both cavities, projecting us into the dual-rail Bell state
\begin{equation}
    \ket{\Phi_{\text{DR}}}=\frac{\ket{1,0,0,1}_{1234}+\ket{0,1,1,0}_{1234}}{\sqrt{2}}
\end{equation}
If we obtain even joint-photon number parity outcomes, we are most likely in the vacuum state of the cavities and should restart the protocol.
\newpage
A step-by-step procedure may look like the following:
\begin{enumerate}
    \item Begin in the state $\ket{1,0}_{12}$
    \item Create $\ket{\text{steady}}\bra{\text{steady}}_{12}$ via dissipation in the bus mode. 
    \item Swap mode 1 and mode 3.  Also swap mode 2 and mode 4
    \item Prepare $\ket{1,0}_{12}$ and use dissipation in the bus mode to again produce $\ket{\text{steady}}\bra{\text{steady}}_{12}$
    \item The state in the cavities should now be $\ket{\text{steady}}\bra{\text{steady}}_{12}\otimes\ket{\text{steady}}\bra{\text{steady}}_{34}$. Measure the joint parity twice on cavity pairs (1,3) and (2,4). These are local measurements performed in each module.
    \item{Only keep attempts where both joint parities are measured to be odd, which should project us into the dual-rail Bell state. Otherwise, restart.}
\end{enumerate}
 
\section{Multiround entanglement generation}
\label{app:mr}
One advantage of using our scheme near critical coupling is that upon failing the vacuum check, the cavity modes will be in vacuum which in principle means the system can be rapidly re-initialized for the next attempt. To investigate how well this quasi-deterministic (repeat until success) entanglement generation performs, we use a different reset sequence after failing the vacuum check, which prioritizes reset speed.
\\
\\
After failing the vacuum check, we are able to reset the system within \SI{5}{\micro\second}. Depending on the outcome $(g,e),(e,g)$ or $(e,e)$, we first reset the transmons to their ground state with the appropriate $\pi$-pulses. Most of the time, the cavities will be in vacuum if we started from a bright state. However, dark states may also fail the vacuum check due to transmons heating out of the ground state and so we always Q-switch both cavities to the bus mode after failed vacuum checks to ensure they are indeed empty. 
\\
\\
This simultaneous Q-switching is performed in \SI{4.5}{\micro\second} by detuning $g_1$ from resonance by \SI{+130}{kHz} and detuning $g_2$ from resonance by \SI{-130}{kHz}. These settings ensure no matter what states are in the cavities, they will be rapidly swapped to the bus. As a consequence of these detunings, the $\phi$ that defines whether a state is bright or dark is changing linearly in time, so all states end up losing energy to the bus eventually.
\\
\\
We find on average 2.6 attempts are required until we successfully pass the vacuum check. This corresponds to an average wait time of \SI{23}{\micro\second} to generate entanglement or an entanglement rate of \SI{43}{kHz}. The measured entanglement fidelity was $86\pm1\%$. For this particular calibration of the experiment, when we use the full cooling sequence (which takes roughly 5 times longer), we obtain an entanglement fidelity of $89\pm1\%$, suggesting our fast reset routine introduces 3\% additional error. This is likely due to transmon reset errors, especially if the transmon has leaked to higher excited states such as the $\ket{f}$ level.  
\section{Pulses for transmon rotations, transmon readout, cavity displacements and parity measurements}
All cavity displacements, including those used in Wigner tomography were Gaussian pulses with a \SI{10}{\nano\second} value of $\sigma$ and a $4\sigma$ duration. 
Transmon $\pi/2$ and $\pi$ pulses were both performed with $\sigma=$\SI{20}{\nano\second} and a $4\sigma$ duration. 
We refer to these as unselective $\pi$ pulses, since the rotation of the transmon state should be independent of the number of cavity photons. 
Number selective transmon pulses used in the vacuum check were also Gaussian pulses with duration $4\sigma$, centered on the transmon $\omega_{ge}$ transition when the cavity was in vacuum. 
For transmon 1 we used $\sigma=$\SI{200}{\nano\second}, and for transmon 2 we used $\sigma=$\SI{300}{\nano\second}, with a longer duration to compensate for the small value of $\chi_{a_2t_2}$.
After a selective $\pi$-pulse, transmon 1 (2) would be read out in its excited state with 94.5\% (96\%) probability, due to transmon dephasing and decay during this rather long pulse. 
\\
\\
For transmon readout on module 1 (2), a \SI{500}{\nano\second} (\SI{700}{\nano\second}) square pulse was used to displace the readout resonator with a total acquisition time of \SI{1.5}{\micro\second}. Single shot readout was achieved via an SPA providing 20dB of gain and additional amplification. (See Fig.\,\ref{fig:wiring}.)
\\
\\
Transmon thermal populations were relatively high, with transmon 1 (2) having an excited population of 4\% (6\%). By using the same feedback cooling routine employed in \cite{Luke2020}, this was reduced to 0.9\% (0.5\%) at the beginning of pulse sequences. This routine amounts to applying $g_1$ and then $g_2$ couplings sequentially to Qswitch any residual excitations in the cavities to the lossy bus mode. We then use selective $\pi$ pulses to verify the cavities are in the vacuum and then finally use measurement-based feedback cooling on the transmons. \\
\\
Cavity thermal populations were 1\% and excited state population became immeasurably small after cooling via measurement. Although we were unable to measure any thermal population in the bus mode, this cooling routine would also ensure the bus mode begins in vacuum. Due to the dispersive coupling of the transmon to both a cavity and the bus mode, we only excite the transmon with the selective $\pi$-pulse when both modes are in vacuum.
\\
\\
\end{document}